\documentclass[useAMS,usenatbib]{mn2e_personal}

\usepackage{graphicx}

\title[Molecular dissipation and macroscopic physical viscosity]{Molecular dissipation in the nonlinear eddy viscosity in the Navier-Stokes equations: modelling of accretion discs}
\author[G. Lanzafame]{G. Lanzafame\thanks{E-mail:
glanzafame@oact.inaf.it}\\
INAF - Osservatorio Astrofisico di Catania, Via S. Sofia
              78 - 95123 Catania, Italy\\}
\begin{document}

\date{Accepted -------. Received -------; in original form -------}

\pagerange{\pageref{firstpage}--\pageref{lastpage}} \pubyear{2009}

\maketitle

\label{firstpage}

\begin{abstract}
Physical damping, regarding the nonlinear Navier-Stokes viscous flow dynamics, refers to a tensorial turbulent dissipation term, attributed to adjacent moving macroscopic flow components. Mutual dissipation among these parts of fluid is described by a braking term in the momentum equation together with a heating term in the energy equation, both responsible of the damping of the momentum variation and of the viscous conversion of mechanical energy into heat. \par
  A macroscopic mixing scale length is currently the only characteristic length needed in the nonlinear modelling of viscous fluid dynamics describing the nonlinear eddy viscosity through the kinematic viscosity coefficient in the viscous stress tensor, without any reference to the chemical composition and to the atomic dimensions. Therefore, in this paper, we write a new formulation for the kinematic viscosity coefficient to the turbulent viscous physical dissipation in the Navier-Stokes equations, where molecular parameters are also included. \par
  Results of 2D tests are shown, where comparisons among flow structures are made on 2D shockless radial viscous transport and on 2D damping of collisional chaotic turbulence. An application to the 3D accretion disc modelling in low mass cataclysmic variables is also discussed. \par
  Consequences of the kinematic viscosity coefficient reformulation in a more strictly physical terms on the thermal conductivity coefficient for dilute gases are also discussed. \par
  The physical nature of the discussion here reported excludes any dependence by the pure mathematical aspect of the numerical modelling.
\end{abstract}


\section{Introduction}

  Physical dissipation in the viscous fluid dynamics is the only physical mechanism for the attenuation of the momentum transfer and for the conversion of mechanical energy (kinetic $+$ potential) into heat. It originates from microscopic particle interactions on molecular scale lengths. For this reason, it is not currently included in the nonlinear Navier-Stokes equations for viscous flows, where macroscopic spatial resolution lengths of moving fluid components are much larger than molecular scale lengths.

  Thus, a physical turbulent viscosity is used in the Navier-Stokes equations, as a tensorial viscous dissipation term, relative to mutual interactions among contiguous macroscopic moving flow parts, producing their braking and their contemporary heating. A turbulent kinematic viscosity coefficient is characterized by a macroscopic scale length multiplied by a scale velocity in the Von K\'arm\'an description, originally formulated to describe a repeating pattern of swirling vortices caused by the unsteady separation of flow of a fluid over bluff bodies. Hence, a mixing length is often required in the formulation of the kinematic viscosity coefficient $\nu$ in the viscous stress tensor describing the nonlinear turbulent eddy viscosity, without any reference to the chemical composition and to the microscopic molecular dimensions. Moreover, physical turbulent viscosity often includes arbitrary parameters, to be set case by case, as it is for of the well known \citet{c50,c51} formulation for disc structures.

  A dissipation mechanism is always necessary in the computational collisional fluid dynamics, even in the non viscous modelling to solve the strictly hyperbolic Euler equations, if flow discontinuities (the Riemann problem) must be solved. In the physically inviscid fluid dynamics, "shock capturing" methods adopt either an artificial viscosity contribution or take advantage of some moderate intrinsic numerical dissipation. Instead, "shock tracking" methods develop a dissipation separately handling shock fronts using appropriate Riemann solver algorithms, through algebraic averages between the two left-right sides as Rankine-Hugoniot jump conditions, in the Godunov-type methods. In addition, a further dissipation is unavoidably intrinsically generated by truncation errors \citep{c3,c4}. In the finite difference methods, dissipation comes about by second order derivatives coming from the Taylor series expansion for incremental ratios of the first order spatial derivatives \citep{a19}, especially for implicit integration techniques. Such a dissipation contribution is also useful to smooth out spurious heating and to treat transport phenomena.

  The physical, turbulent dissipation and the fictitious artificial or numerical ones are conceptually distinct, although formally similar, as shown in \citet{c48,a9,a11}, and discussed by \citet{c45,c46} in the case of smooth particle hydrodynamics (SPH) accretion disc modelling in close binaries (CBs), or by \citet{a71} in the case of finite difference or for finite volume integration techniques. In both cases, some arbitrary parameters, to be tuned case by case, are included and/or a dependence on the spatial resolution length affects the numerical dissipation.

  Therefore, in this paper we propose the formulation of a physical turbulent kinematic viscosity coefficient in the Navier-Stokes equations, free of any arbitrary parameters, where microscopic physical characteristics are also included. A macroscopic scale length (mixing length) is obviously still used, since solutions of the nonlinear Navier-Stokes equations, involve macroscopic physical properties. At the same time, a reformulation for $\nu$ in a more physical sense is also coherent to a reformulation for the thermal conductivity coefficient $c$ for dilute gases.

  In \S 2 of this paper we discuss some general aspects of dissipation in the computational collisional fluid dynamics; in \S 3, we shortly describe characteristics of some adopted physical turbulent kinematic viscosity coefficients $\nu$, while in \S 4 we formulate the physical development of both $\nu$ and $c$ coefficients including microscopic molecular characters. In \S 5 we show some results for some essential 2D tests on shockless radial viscous transport in an annular ring, as well as for the damping of 2D Burger's turbulence. Instead, in \S 6, we show an astrophysical application in the case of a 3D accretion disc modelling in a low mass close binary (LMCB) system, comparing 3D accretion disc structures obtained by using different $\nu$ coefficients and gas compressibility.

  Despite the adoption of a numerical technique intrinsically quite viscous, like the SPH, successful physically viscous results unconditionally show that the new formulation for the kinematic viscosity coefficient works well without any restriction.

\section{Dissipation in viscous and non viscous fluid dynamics}

  In the physically non viscous flows, the hyperbolic Euler system of equations
  
\begin{equation}
\frac{d\rho}{dt} + \rho \nabla \cdot \bmath{v} = 0 \hfill \mbox{continuity equation}
\end{equation}

\begin{equation}
\frac{d \bmath{v}}{dt} = - \frac{\nabla p}{\rho} + \bmath{f} \hfill \mbox{momentum equation}
\end{equation}

\begin{equation}
\frac{d}{dt} \left( \epsilon + \frac{1}{2} v^{2}\right) = - \frac{1}{\rho} \nabla \cdot \left( p \bmath{v} \right) + \bmath{f} \cdot \bmath{v} \hfill \mbox{energy equation}
\end{equation}

\begin{equation}
\frac{d \bmath{r}}{dt} = \bmath{v} \hfill \mbox{kinematic equation.}
\end{equation}

  must be solved, together with the state equation (EoS) of the fluid
  
\begin{equation}
p = f(\gamma, \rho, \epsilon, \bmath{r}, \bmath{v}) \hfill \mbox{state equation}
\end{equation}
  
  Most of the adopted symbols have the usual meaning: $d/dt$ stands for the Lagrangian derivative, $\rho$ is the gas density, $\epsilon$ is the thermal energy per unit mass, $p$ is the ideal gas pressure, here generally expressed as a function of local properties, $\bmath{v}$ and  $\bmath{r}$ are the vector velocity and position, $\bmath{f}$ is the external force field per unit mass. The adiabatic index $\gamma$ has the meaning of a numerical parameter whose value lies in the range between $1$ and $5/3$.

  Since the Riemann problem must be correctly solved for collisional flows in the case of shocks, a dissipation mechanism is necessary otherwise frontal colliding flows trespass each other. Such a dissipation could be either explicit, as an artificial viscosity term for shock capturing schemes, especially for finite volume schemes, or it could be intrinsic through a specific Riemann solver code \citep{c37,c3,c4,c5,a19} either for shock tracking schemes or for Eulerian finite difference schemes by commuting mathematical derivatives in incremental ratios. As an example, in the finite difference techniques, the conversion of the 1st order spatial derivative of the generic physical quantity $u$ is $(u_{i} - u_{i-1})/\Delta x$, where $i$ is a spatial grid index and $\Delta x$ is the grid spatial resolution length. For a better stable result, the same incremental ratio is rewritten as

\begin{equation}
\frac{u_{i} - u_{i-1}}{\Delta x} = \frac{u_{i+1} - u_{i-1}}{2 \Delta x} - \frac{u_{i+1} - 2u_{i} + u_{i-1}}{2 \Delta x},
\end{equation}

which gives a much better stability, at the cost of a reduction in accuracy. The contribution of the second term $(u_{i+1} - 2u_{i} + u_{i-1})/\Delta x$ analytically corresponds to a 2nd order spatial derivative, working exactly like a real viscous non physical contribution. These manipulations of spatial derivatives in the incremental ratios in finite terms are necessary to ensure stability to the solutions of hyperbolic systems of equations. This means that the inclusion, or the numerical development of a non physical dissipation distorts numerical results especially for non collisional events like shear flows or transport phenomena. These numerical difficulties arise when the EoS
  
\begin{equation}
p = (\gamma - 1) \rho \epsilon \hfill \mbox{perfect gas equation}
\end{equation}

is adopted for ideal flows. Instead, an EoS as:

\begin{equation}
p^{\ast} = \frac{\rho}{\gamma} c_{s}^{2} \left(1 - C \frac{n^{-1/3} \nabla \cdot \bmath{v}}{3 c_{s}} \right)^{2},
\end{equation}

includes a real macroscopic physical dissipation correctly handling the Riemann problem, as well as transport and shear flows free of any local gas compression \citep{c16,c24,c25}. $c_{s}$ is the sound velocity, $n$ is the numerical density, while

\begin{equation}
C = \frac{1}{\pi}  \textrm{arccot} \left( D \frac{v_{R}}{c_{s}} \right),
\end{equation}

where $D \gg 1$ and where $v_{R}$ is the component of velocity along the direction of collision. $D$ is a large number describing how much the flow description corresponds to that of an ideal gas: $D \approx \lambda/d$, being $\lambda \propto \rho^{-1/3}$ the molecular mean free path, and being $d$ the mean linear dimension of gas molecules. The physical dissipation, expressed by the two further terms in eq. (8) (the linear and the quadratic terms in $\nabla \cdot \bmath{v}$) of the reformulated EoS, better treats both shocks and shear flows, even in a Lagrangian description. Their inclusion substitutes artificial viscosity terms and does not represent a physical turbulent viscous contribution, but the real physical dissipation coming out because eq. (7) of the EoS should strictly be applied only to macroscopic static or quasi-static processes. Notice that in this case, this real macroscopic physical dissipation does not originate from a physical viscosity. Instead, it originates from the irreversible thermodynamic process and is better evidenced in a Lagrangian description.

  In the physically viscous flows, the Navier-Stokes equations explicitly include macroscopic physical dissipation terms in the momentum and in the energy equations:

\begin{eqnarray}
\frac{d \bmath{v}}{dt} & = & - \frac{\nabla p}{\rho} + \bmath{f} + \frac{1}{\rho} \nabla \cdot \bmath{\tau} \nonumber \\
& & \ \ \ \ \ \ \ \ \ \ \ \ \ \ \ \hfill \mbox{Navier-Stokes momentum equation}
\end{eqnarray}

\begin{eqnarray}
\frac{d}{dt} \left( \epsilon + \frac{1}{2} v^{2}\right) = - \frac{1}{\rho} \nabla \cdot \left[ \left( p \bmath{v} - \bmath{v} \cdot \bmath{\tau}) + c \nabla (\rho \epsilon \right) \right] + \bmath{f} \cdot \bmath{v} \nonumber
\end{eqnarray}

\begin{equation}
\hfill \mbox{Navier-Stokes energy equation,}
\end{equation}

  where the viscous stress tensor $\bmath{\tau}$ and the thermal conductivity $1/\rho \nabla \cdot [c \nabla (\rho \epsilon)]$ terms are explicitly added, to be solved together with the continuity equation, the kinematic equation and the EoS. It is important to note that the thermal flux term includes two contribution: the first contribution depends on the thermal gradient ($\nabla \epsilon$) that is currently used for solids or for incompressible fluids, while the second one depends on the density gradient ($\nabla \rho$) correlated to the mass diffusivity, here not included in the continuity equation. The matrix element $(\alpha, \beta)$ of the viscous stress tensor

\begin{equation}
\tau_{\alpha, \beta} = \eta \sigma_{\alpha, \beta} + \zeta \nabla \cdot \bmath{v}
\end{equation}

and

\begin{equation}
\sigma_{\alpha, \beta} = \frac{\partial v_{\alpha}}{\partial x_{\beta}} + \frac{\partial v_{\beta}}{\partial x_{\alpha}} - \frac{2}{3} \delta_{\alpha, \beta} \nabla \cdot \bmath{v}.
\end{equation}

  $\eta$ and $\zeta$ are the dynamic first (shear) and second (bulk) physical viscosity coefficients.
  
  In the present study, we simply consider $\zeta = 0$ and eq. (7) as EoS. By definition, the physical kinematic viscosity coefficient is $\nu = \eta/\rho$.

\section{The physical kinematic viscosity coefficient}

  Typical kinematic laboratory viscosities are of the order of $\nu = 0.001 - 1$ cm$^{2}$ s$^{-1}$, to be compared with inertial forces in the ratio

\begin{equation}
Re = \frac{\mbox{inertial forces}}{\mbox{viscous forces}} \equiv \frac{l_{flow} v_{flow}}{\nu}
\end{equation}

where $l_{flow}$ and $v_{flow}$ are the characteristic length and velocity scales of the microscopic flow. Laboratory experience shows that for $Re > Re_{crit} \approx 10^{2} - 10^{3}$, flow becomes turbulent. $Re_{crit}$ is the characteristic Reynolds number as observed so far.

  In the full nonlinear approach, the full non linearity of the Navier-Stokes equations is considered, where spatial derivatives of the entire velocity field are used. Neither the Reynolds averages of the Navier-Stokes equations in boxes of intermediate size (as in the linear approach), nor the full Navier-Stokes equations working with spatial gradients of the mean velocity field (as in the nonlinear Boussinesq approach \citep{b10}), are considered.

  To characterize a nonlinear macroscopic physical kinematic viscosity coefficient $\nu$, characteristic length and velocity scales $l$ and $v$ are needed, which are unknown, in principle.

  Typically \citep{b1}, a mixing length model can be used, where

\begin{equation}
v \sim l \left| \frac{\partial v}{\partial x} \right|,
\end{equation}

\begin{equation}
\nu \sim l^{2} \left| \frac{\partial v}{\partial x} \right|,
\end{equation}

or, more generally, for a better statistical evaluation,

\begin{eqnarray}
\frac{\nu^{2}}{l^{4}} & \sim & \left( \frac{\partial v_{x}}{\partial y} + \frac{\partial v_{x}}{\partial z} \right)^{2} + \left( \frac{\partial v_{y}}{\partial x} + \frac{\partial v_{y}}{\partial z} \right)^{2} + \nonumber \\ & & \left( \frac{\partial v_{z}}{\partial x} + \frac{\partial v_{z}}{\partial y} \right)^{2} .
\end{eqnarray}

Here, the problem relies in the evaluation of $l$. Being $h$ the computational spatial resolution, and being $L$ the scale length of the entire computational domain, $h \leq l \leq L$. Because of the lack of any geometric information, the only physical scale lengths we know are those relative to the hydrostatic equilibrium (in the presence of an external force field): $\int dp/\rho f$, as well as $p/|\nabla p|$, $\rho/|\nabla \rho|$, $|\bmath{v}|/\nabla \cdot \bmath{v}$, $|\bmath{v}|/|\nabla \times \bmath{v}|$, etc.. 

  Since, in 3D the natural tendency is the development of smaller structures in a direct cascade process \citep{a72,a73}, some authors \citep{b2} calculate

\begin{equation}
l = \left( \sum_{i} l_{i}^{-1}  \right)^{-1}.
\end{equation}

where $l_{i}$ refers to various scale lengths as $(\partial \ln \rho/\partial r)^{-1}$, $(\partial \ln V/\partial r)^{-1}$, $(\partial \ln p/\partial r)^{-1}$, $(\nabla \cdot \bmath{v}/|\bmath{v}|)^{-1}$, etc..

  Instead, no information we have about $v$, since we only know $v_{flow}$ and $c_{s}$.
  
  In the viscous accretion disc modelling, the Shakura and Sunyaev \citet{c50,c51} parametrization of turbulent viscosity is largely adopted. In this approach, the kinematic viscosity coefficient is

\begin{equation}
\nu = \frac{1}{3} l v,
\end{equation}

where both $l$ and $v$ are unknown. Assuming the flow isotropy, a Keplerian tangential kinematics, and the vertical hydrostatic equilibrium,

\begin{equation}
l = \alpha_{l} H,
\end{equation}

where $H \simeq r c_{s}/v_{Kepl}$, the local disc thickness, is the local shortest macroscopic scale length, and $\alpha_{l} \leq 1$ is a scaling quantity. Without any isotropy assumption, $\alpha_{l} > 1$. At the same time,

\begin{equation}
v = \alpha_{v} c_{s},
\end{equation}

where $\alpha_{v} \leq 1$. Whenever $v > c_{s}$, shocks would dissipate the energy, reducing the velocity to subsonic. Hence, in the Shakura and Sunyaev approach,

\begin{equation}
\nu = \frac{1}{3} \alpha_{l} \alpha_{v} c_{s} H = \alpha_{SS} c_{s} H,
\end{equation}

with $\alpha_{SS} < 1$ to be found.

  \citet{b3} found $0.01 \leq \alpha_{SS} \leq 0.03$ for $0.1 \leq {\mathcal L/\mathcal L}_{E} \leq 1$ (${\mathcal L}_{E} = $ Eddington luminosity) as a lower limit for active galactic nuclei (AGN). For numerical simulations of AGN, $\alpha_{SS} \approx 10^{-4} - 10^{-3}$ is often adopted \citep{b22,b23}. Values for $\alpha_{SS} \approx 10^{-2}$ have also been found for the observed protostellar objects \citep{b4}, while for a fit of FU Orionis observed outburst \citep{b5,b6,b7}, $\alpha_{SS} \approx 0.001 - 0.003$. In fully ionized discs in dwarf novae, the best observed evidence suggests a typical $\alpha_{SS} \sim 0.1 - 0.4$, whilst the relevant numerical MHD simulations evaluate $\alpha_{SS} \approx 10$ times smaller. \citet{c52} attribute this discrepancy to incorrect magnetic and boundary layer shortcomings in the computations. Nevertheless, this is not the conclusion of the full story because \citet{c45,c46} showed that a well bound viscous accretion disc structure modelling strongly depends on several conditions: the kinematic of the mass transfer, $\gamma$, $\alpha_{SS}$ and so on. For isothermal or for a quasi-isothermal thermodynamics, a disc is structurally bound even for $\alpha_{SS} = 0$ because numerical dissipation only is enough to produce a disc in shocks events, a result also discussed by \citet{b8,b9} and, more recently, by \citet{c16} in its physical sense.

\section{$\nu$ and $c$ coefficients and molecular characteristics}

  Different formulations of macroscopic physical dissipation rarely converge with each other and often include an arbitrary parameter, to be evaluated. In addition, any correlation to microscopic (molecular, atomic, nuclear) physical properties is absent.

  To this purpose, we propose the following evaluation of the physical kinematic viscosity coefficient $\nu$, to be used to determine the viscous stress tensor $\bmath{\tau}$ in the Navier-Stokes equations.

  Microscopic molecules, atoms, nuclei, have known elastic scattering impact cross section $\kappa$, useful to compute $\nu$. Without any consideration to the existence or to the involvement of internal energy levels for an ideal flow, physical dissipation transfers macroscopic ordered kinetic energy flows into heat, that is in microscopic chaotic kinetic energy flows. This means that elastic scattering collisional cross sections for an ideal gas are those right for a reformulation of $\nu$. For a gas mixture, the mean value of the elastic scattering impact cross section

\begin{equation}
\overline{\kappa} = \sum_{i} X_{i} \kappa_{i}
\end{equation}

should be considered, where $X_{i} = n_{i}/\sum_{i} n_{i}$ is the relative numerical abundance of the chemical species $i$.

  However, we need to take into account the total number of microscopic molecules-atoms within the macroscopic cross section determined by the mixing length as in eq. (18) which arithmetically gives as a result a mixing length close to the smallest scale length in the summation. For this reason, we prefer to compute $l$ as

\begin{equation}
l = \mbox{min} (l_{1}, l_{2}, l_{3}, ..., l_{n}),
\end{equation}

the various $l_{i}$ referring to $p/|\nabla p|$, $\rho/|\nabla \rho|$, $|\bmath{v}|/\nabla \cdot \bmath{v}$, $|\bmath{v}|/|\nabla \times \bmath{v}|$, and so on.

  Hence, the total number of barions contained within the 3D mixing mass $\rho l^{3}$ is $\rho l^{3} /\overline{\mu} m_{H}$, where $m_{H}$ is the proton mass and $\overline{\mu}$ is the mean molecular weight.

  Since we need the molecular collisional cross section, in 3D it is necessary to compute $(\rho l^{3}/\overline{\mu} m_{H})^{2/3}$, that is the number of molecules within the volume $l^{3}$, powered to $2/3$. This number has to be multiplied with $\overline{\kappa}$ to get a statistically effective collisional surface composed of a multitude of microscopic cross sections:

\begin{equation}
\left( \frac{\rho l^{3}}{\overline{\mu} m_{H}} \right)^{2/3} \overline{\kappa}
\end{equation}

  To calculate $\nu$, we need to divide this arithmetic term by a length $\lambda$. This length decreases whenever the local numerical density increases. Therefore $\lambda \sim n^{-1/3}$.

  As far as the velocity contribution of $\nu$ is concerned, we exclude not only $v_{flow}$, but also the thermal $\epsilon^{1/2}$, being this last strictly linked to the chaotic microscopic kinematics. Kinematic velocities of extraneous bodies - as in the original Von K\'arm\'an formulation - moving in the fluid are not considered. This exclusion is a consequence of the fact that the kinematic viscosity coefficient, mutually correlated to the thermal conductivity and the diffusivity coefficients, are all expressions of the intrinsic physical property of the fluid. Hence, to conclude, our

\begin{equation}
\nu \simeq \xi = \left( \frac{\rho l^{3}}{\overline{\mu} m_{H}} \right)^{2/3} \overline{\kappa} n^{1/3} c_{s} = \frac{\rho l^{2}}{\overline{\mu} m_{H}} \overline{\kappa} c_{s},
\end{equation}

being $n = \rho/\overline{\mu} m_{H}$.

  Its 2D counterpart, according to the same algebraic logical steps is:
  
\begin{equation}
\nu \simeq \xi = \left( \frac{\Sigma l^{2}}{\overline{\mu} m_{H}} \right)^{1/2} \overline{\kappa} n^{1/2} c_{s} = \frac{\Sigma l}{\overline{\mu} m_{H}} \overline{\kappa} c_{s},
\end{equation}

being $\Sigma$ the 2D mass density.

  In these expressions, both the molecular/atomic $\overline{\mu} m_{H}$ and $\overline{\kappa}$ are included, as well as other macroscopic physical quantities like $\Sigma$ or $\rho$. This means that, the spatial component of $\nu$ could strictly be not shorter than $l$ and longer than $h$, according to other physical variables, especially $\Sigma$ or $\rho$. This formulation is free of any arbitrary parameter. Moreover, in (26) and (27), the quantities $(\rho/\overline{\mu} m_{H}) l \overline{\kappa}$ and $(\Sigma/\overline{\mu} m_{H}) \overline{\kappa}$ are pure numbers both $< 1$, otherwise the density of the fluid is comparable or greater than the atomic density. It is important to note that eqs. 26 and 27 are not exactly equivalent. In eq. 26 $\nu \propto l^{2}$ in its spatial dependence (as Prandtl's eqs. 16, 17), while in eq. 27 $\nu \propto l$ in the same spatial contribution (as Shakura and Sunyaev's eq. 22). Eqs. 26 and 27 should be considered strictly correlated either to a 3D or to a 2D modelling, respectively. Indeed, densities $\rho$ and $\Sigma$ are formally correlated by the equivalence of their numerical densities as $(\Sigma/\overline{\mu} m_{H})^{3} = (\rho/\overline{\mu} m_{H})^{2}$. So that $\Sigma \equiv \rho l$ would be a very specific case.

  Molecular or atomic collisional cross sections are assumed, for the sake of simplicity, circular, without any distortion. Molecular dimensions are determined by the so called Van der Waals mean radius, defining the limit where the microscopic molecular force potential becomes attractive, deviating from that of a non interacting free particle relative to an ideal gas. In the case of fully ionized gas, the repulsive Coulomb elastic scattering among head on colliding ions determines the shortest classical impact parameter $r_{p}$ as $r_{p} \simeq (4 \pi \epsilon_{\circ})^{-1} 2 Z_{1} Z_{2} e^{2}/3 K_{B} T$, related to $\overline{\kappa}$ as $\overline{\kappa} = \pi r_{p}^{2}$. $Z_{1} e$ and $Z_{2} e$ are the two effective electric charges of the two colliding ions, $K_{B}$ is the Boltzman constant, $T$ is the temperature and $\epsilon_{\circ}$ is the dielectric constant of the vacuum.

  Notice that the decrease of the mixing length $l$ in a more effective characteristic scale length as a result of the presence of a multitude of small scale constituents, as shown in eqs. (26, 27), does not alter the meaning of the kinematic viscosity coefficient role in the tensorial expression of the stress viscous tensor (eqs. 12, 13) in the Navier-Stokes equations describing fluid flows. Therefore, although we introduce local physical properties of the fluid, and although turbulence is not a feature of fluids but of fluid flows, the reformulation of the scalar $\nu$ within $\tau_{\alpha, \beta}$ stays meaningful.

\subsection{Kinematic viscosity and thermal conductivity coefficients for dilute gases}

  The kinematic viscosity coefficient $\nu$ and the thermal conductivity coefficient $c$ are dimensionally identical. Both are characterized by a scale length multiplied by a scale velocity. Hence, both are transport coefficients. A relevant difference between viscosity and thermal conductivity is that while viscosity is activated only whenever a relative motion occurs among contiguous flow elements, thermal conductivity is an energy transport mechanism always existing whenever a temperature gradient occurs even in steady state conditions. However, the two coefficients are always related to each other because both explains the tendency of the thermodynamic system toward a homogeneity and isotropy status, smoothing out kinematic ($\nu$) and thermal ($c$) local spatial discrepancies. The ratio $c/\nu$ \citep{a39} is:

\begin{equation}
\frac{c}{\nu} = \frac{c_{V}}{\epsilon \overline{\mu}},
\end{equation}

where $c_{V}$ is the molar specific heat of the gas at constant volume which, for an ideal (monoatomic) gas is $c_{V} = 3RT/2 = 3K_{B}T/m_{H}$. Experimentally $(c/\nu) (\epsilon \overline{\mu}/c_{V}) \approx 1.3 - 2.5$, instead of $1$. The discrepancy is largely explained because of the fact that theoretically $c$ is evaluated considering a uniform molecular velocity distribution instead of local molecular kinematic differences related to the presence of a temperature spatial gradient. As a consequence, considering $\nu = \xi$, for an ideal gas,

\begin{equation}
c = 2 \nu \frac{c_{V}}{\epsilon \overline{\mu}} \simeq 2 \xi \frac{c_{V}}{\epsilon \overline{\mu}} = 3 \xi \frac{K_{B} T}{\epsilon \overline{\mu} m_{H}}
\end{equation}

so that,

\begin{equation}
c \simeq 2 \left( \frac{l}{\overline{\mu}} \right)^{2} \rho \frac{\overline{\kappa}}{m_{H}} c_{s} \frac{c_{V}}{\epsilon} = 3 K_{B} T \left( \frac{l}{\overline{\mu} m_{H}} \right)^{2} \rho \overline{\kappa} \frac{c_{s}}{\epsilon}
\end{equation}

in 3D, and

\begin{equation}
c \simeq 2 \left( \frac{1}{\overline{\mu}} \right)^{2} \Sigma l \frac{\overline{\kappa}}{m_{H}} c_{s} \frac{c_{V}}{\epsilon} = 3 K_{B} T \left( \frac{1}{\overline{\mu} m_{H}} \right)^{2} \Sigma l \overline{\kappa} \frac{c_{s}}{\epsilon}
\end{equation}

in 2D. Since coefficients $\nu$ and $c$ are comparable for dilute gases, then the viscous and the thermal conductivity time scales are also comparable. This means that the inertia of matter tends toward a uniform kinematic and thermal configuration with the same time scales.

  In the viscous computational fluid dynamics, currently $\nu$ and $c$ are not only arbitrarily parametrized, but also not correlated from each other, as it correctly should be. In spite of the fact that results in isothermal or in nearly isothermal conditions could still be significant, the lack of any correlation between $\nu$ and $c$ is free of any physical meaning.

  In the rest of the paper, we mainly pay attention to the physical viscosity. However, since now onwards, any conclusion referring to the role of the physical viscosity will also refer to the role of the thermal conductivity in a close cause-effect correlation, where any any significant local spatial derivative involving a relative motion among contiguous flow parts will involve a a braking and a viscous heating; any local heating will involve larger pressure and temperature gradients and consequently a transport of energy and mass with comparable time scales; a heat transfer will involve a decrease of temperature and pressure spatial gradients.

\section{Viscosity tests}

  Flow transport and damping of turbulence are the only two fields where the role of physical dissipation in the Navier-Stokes equations is better emphasized. Therefore, in this section, appropriate tests on the physical viscosity efficiency will be done. In particular, regard a 2D SPH shockless modelling on the radial flow viscous transport in an annulus ring and a 2D SPH modelling of Burger's turbulence. $\gamma = 5/3$ will be assumed throughout the tests. This low compressibility regime is the less advantageous condition for viscous forces against pressure forces, as well as against numerical-artificial damping \citep{c48,a9,a11}, explained by a modest contribution of the bulk component $\nabla \cdot \bmath{v}$ in eqs. (12, 13), not compensated by the high $c_{s}$, on $\nu$. In such tests, results on $\nu = \xi$, also including a thermal conductivity $c \propto \nu$ will also be compared with those relative to

\begin{equation}
\nu = c_{s} h
\end{equation}

\begin{equation}
\nu = c_{s} l
\end{equation}

which are the two simplest analytical expressions for $\nu$, where the characteristic length could be either at its minimum geometric value $h$, or coincident with the mixing length $l$. Despite their full mathematical meaning, formulations (32) and (33), like (eqs. 16, 17, 22), lack of any physical sense without any correlation to what matter the flow is made of.

\subsection{2D radial viscous transport in a shockless isothermal annulus ring}

  Theory on 2D shockless radial transport in a Keplerian annulus ring in a gravitational potential well \citep{b3} predicts that, from the Green function, the solution of the initial Keplerian mass distribution at time $t = 0$ for $\Sigma$ is:

\begin{equation}
\Sigma(r,t=0) = m \delta(r - r_{\circ})/2\pi r_{\circ}
\end{equation}

for a ring having mass $m$ and an initial radius $r_{\circ}$. The solution, at time $t$, in terms of dimensionless radius $x = r/r_{\circ}$ and viscous time $\theta = 12 \nu t r_{\circ}^{-2}$ is

\begin{equation}
\Sigma(x,t) = (m/\pi r_{\circ}^{-2}) \theta^{-1} x^{-1/4} \exp[-(1 + x^{2})/\theta] I_{1/4}(2x/\theta),
\end{equation}

where $I_{1/4}$ is the modified Bessel function. The action of viscosity is to spread out the entire annulus ring toward a disc structure transporting most of the low angular momentum mass toward the centre of the potential well and transporting a smaller fraction of high angular momentum mass toward the empty external space.

  For practical computational purposes, since it is impossible to reproduce a delta Dirac function at time $t = 0$, it is necessary to start numerical calculations from an initial mass distribution relative to a small $\theta$ value. In our example, we choose $\theta = 0.017$, a value comparable with that used by \citet{b11} for a radial viscous transport similar test.

\begin{figure}
\resizebox{\hsize}{!}{\includegraphics[clip=true]{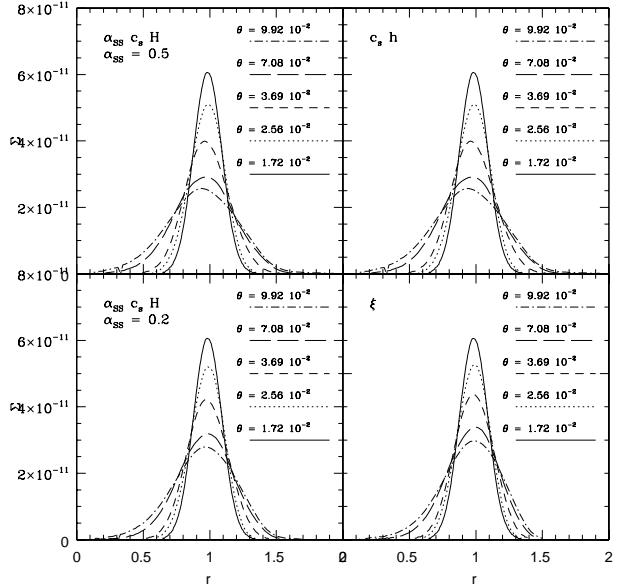}}
\caption{Examples of four calculated mass radial distributions as a function of the viscous time $\theta$. Two $\alpha$ Shakura and Sunyaev radial distributions are shown, as well as a $\nu = c_{s} h$, with $h = 0.09$, compared with our $\nu = \xi$ radial profile.}
\end{figure}

  Fig. 1 shows a comparison among several radial distributions of the 2D mass density $\Sigma(r, \theta)$ for various $\theta$. Because of the universality of graphical representation by using $\theta$ instead of $t$, computational radial profiles at an assigned $\theta$ value have to correspond with each other, fitting the analytical solution (eq. 35). In particular, fig. 1 shows in the same picture $\Sigma(r, \theta)$ profiles both for two Shakura and Sunyaev kinematic viscosity $\nu = \alpha_{SS} c_{s} H$, and for the $\nu = c_{s} h$, where $h = 0.09$, and for the newest $\nu = \xi$, where our formulation for $\nu$ is used. This clearly shows that the viscous $\nu = \xi$ model correctly replicates the right radial shockless transport mechanism, without any local distortion. In this example, we used throughout the models, $M_{\circ} = 2 \cdot 10^{33}$ g, $R_{\circ} = 10^{11}$ cm, as typical astrophysical values. We considered an initial density $\Sigma_{\circ}$ of the order of $10^{-10}$ g cm$^{-2}$, an initial sound velocity $c_{s} = 5 \cdot 10^{-2} v_{\circ}$, where $v_{\circ} = 2 \pi (GM_{\circ}/R_{\circ})^{1/2}$ cm s$^{-1}$ is the normalization value for the velocity, and a gas composed of pure molecular hydrogen. The thermal energy per unit mass $\epsilon$ is kept constant, being $d \epsilon/dt = 0$ throughout the entire simulation. Variations of these parameters do not produce any difference in the radial density distribution for each $\theta$, being the viscous time $\theta$ an absolute reference time for $\Sigma$. The $\nu = c_{s} h$ profile is shown as a further test on the radial profile, where eq. (32) has been considered in the $\nu$ calculation. Instead, models adopting eq. (33), or even the Prandtl formulation (15), do not yield any realistic radial profile of mass distribution because of the too large viscosity, due to the very negligible spatial derivatives in the $l$ calculation, causing the rapid accretion of the entire disc.

  In spite of the low spatial resolution adopted for practical purposes, results on radial shockless viscous transport here reported clearly show that, without any other physical alternative, the Shakura and Sunyaev formulation looks like an appropriate expression for $\nu$ for 2D shockless disc structures, since it uses a characteristic length $H$ much shorter than $l$ along a 3rd dimension that does not exist in 2D. However, physical formulations as in eq. (26) or (27), where the effective scale length $(\Sigma l/\overline{\mu} m_{H}) \overline{\kappa}$ is much shorter than $l$ in the case of diffuse matter, could be a valid physical alternative, free of any arbitrary parameter.

\subsection{Damping of 2D Burger's turbulent flow}

  Statistical studies of turbulence normally involve hypotheses about homogeneity and isotropy on the distribution and kinematics of the spatial flow \citep{a72,a73}. 2D turbulence is relevant to understand large scale flows \citep{a58,a59}. 2D turbulence schematically discusses either a "forced steady state turbulence", or a "decaying turbulence", if an explicit forcing term is added in the momentum equation or not. Eddies of different sizes showing density and potential fluctuations in the flow normally characterize turbulence. 

\begin{figure}
\resizebox{\hsize}{!}{\includegraphics[clip=true]{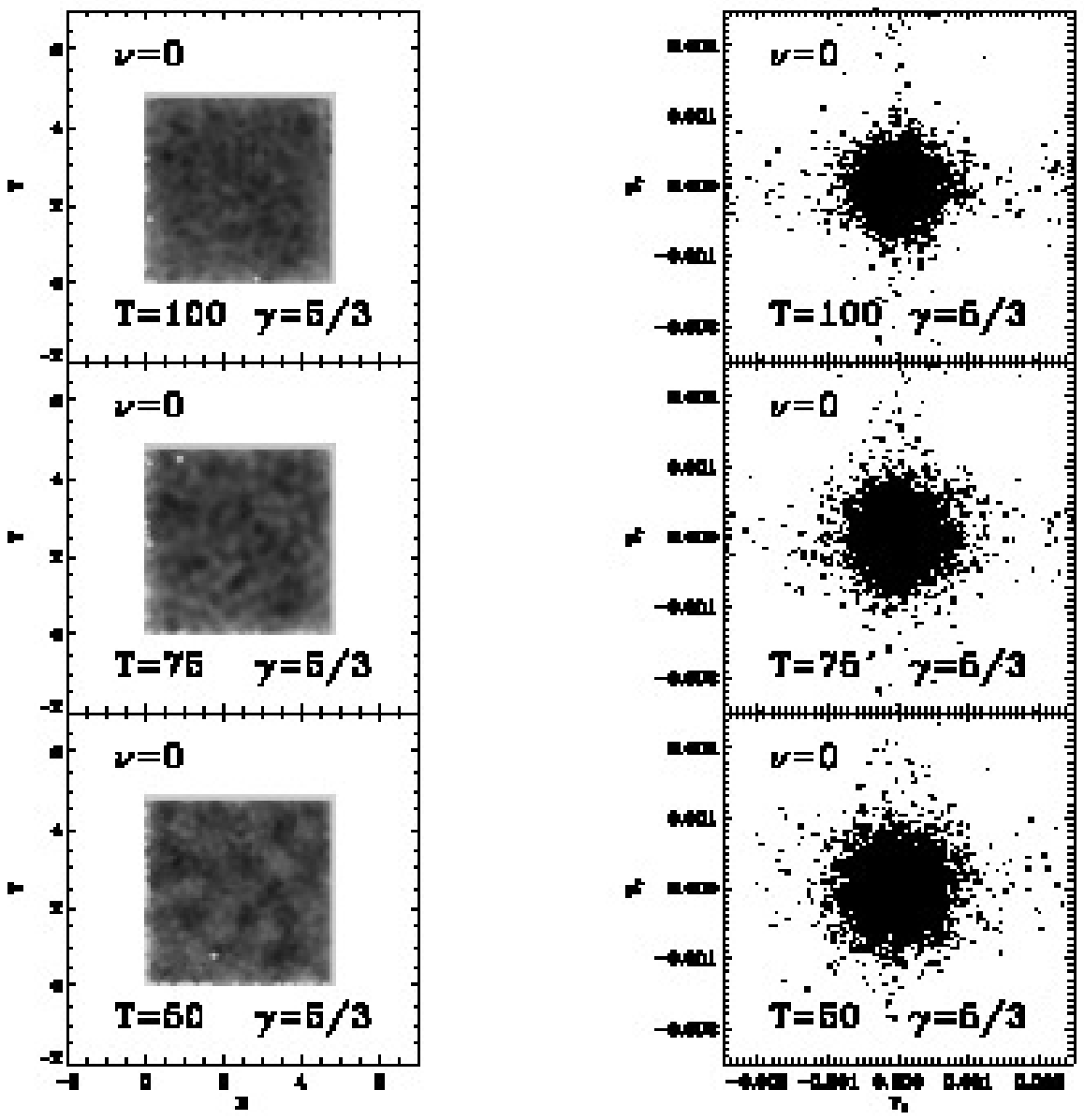}}
\resizebox{\hsize}{!}{\includegraphics[clip=true]{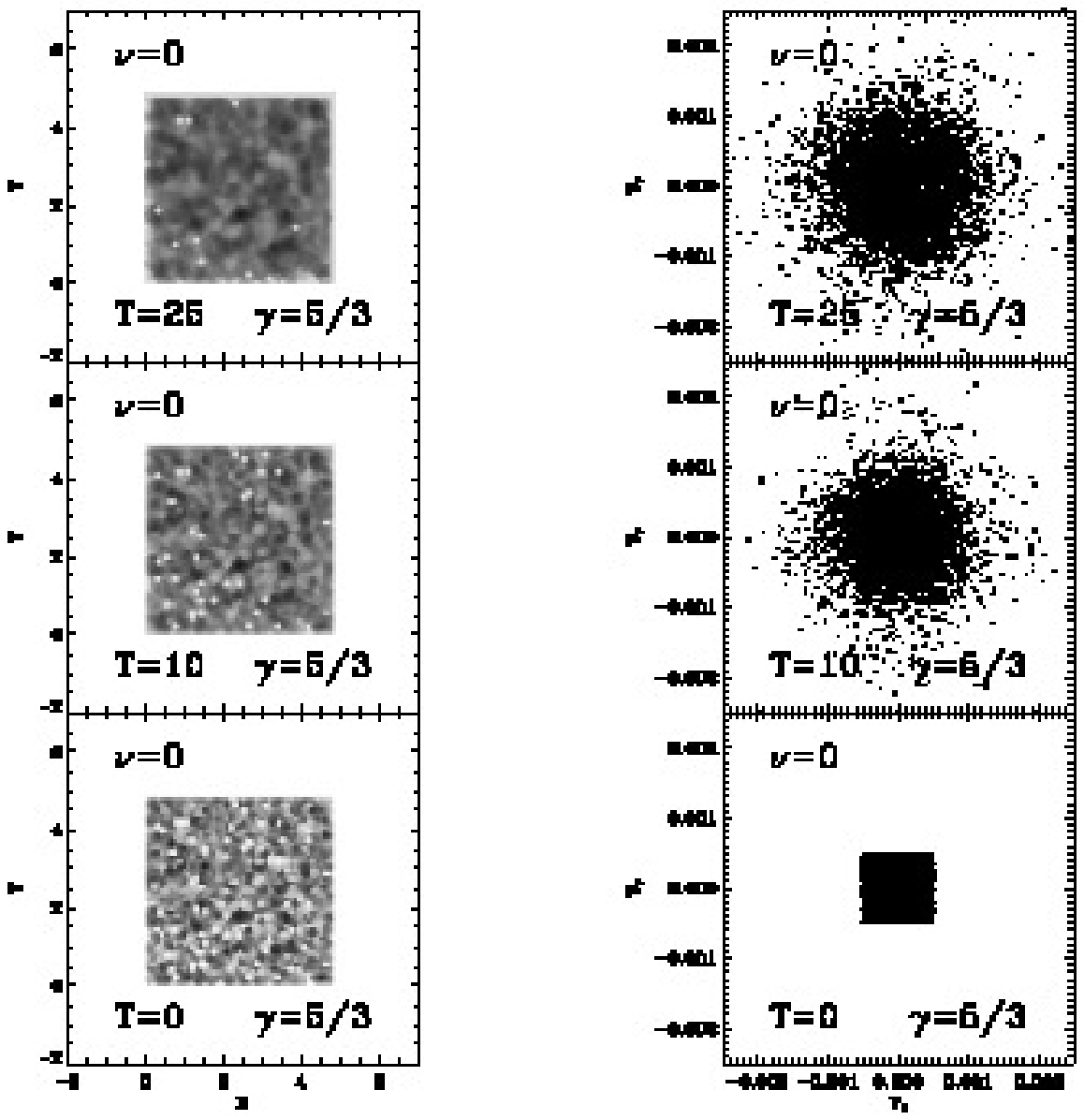}}
\caption{($X,Y$) plots of density map at various times $T$ for the physically non viscous model with $h = 0.05$. $64$ greytone are used. $v_{X}$, $v_{Y}$ tomograms are also reported, showing velocity fluctuation both during the initial turbulent phase and during the following damping subsequent phase.}
\end{figure}

\begin{figure}
\resizebox{\hsize}{!}{\includegraphics[clip=true]{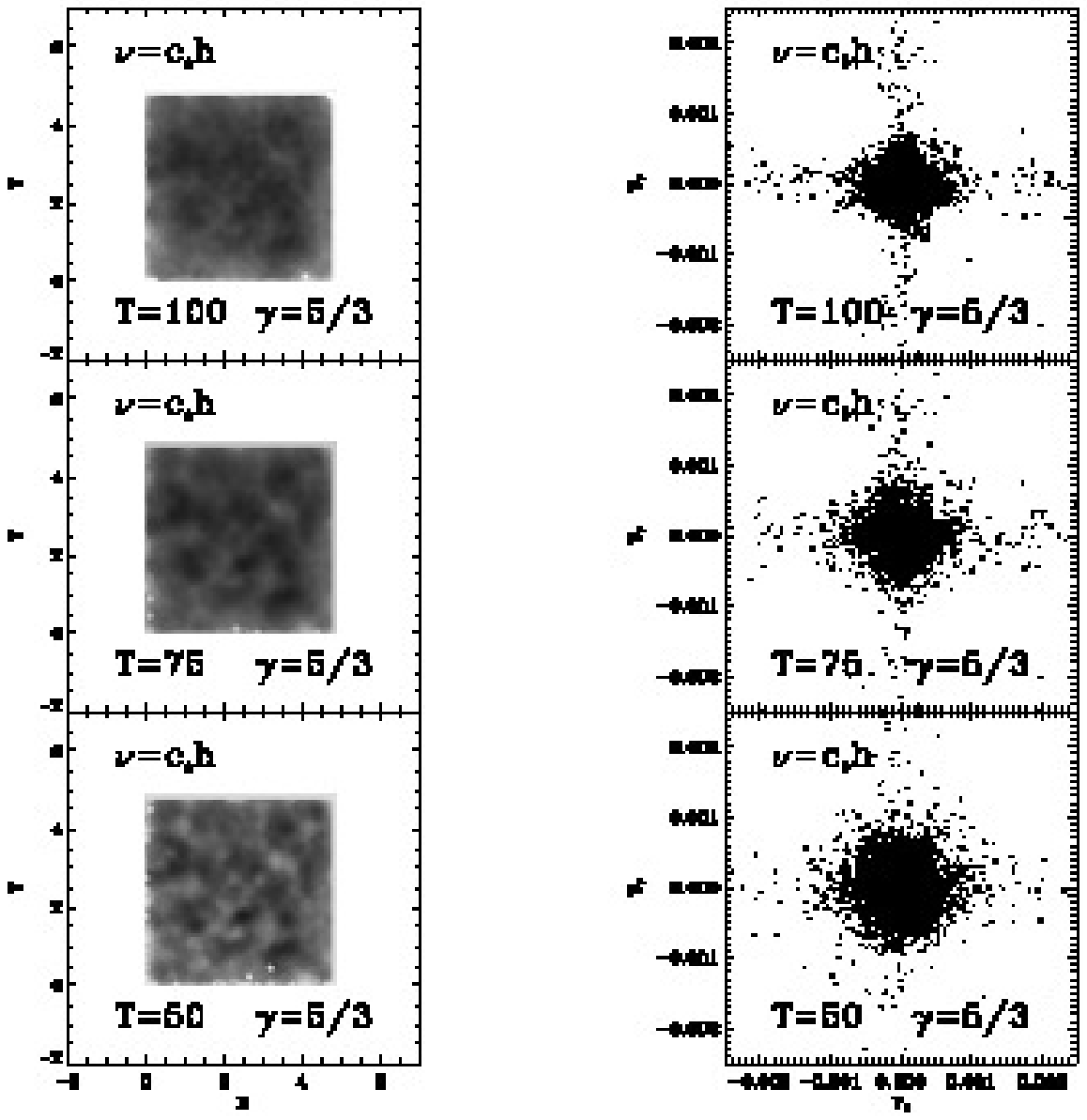}}
\resizebox{\hsize}{!}{\includegraphics[clip=true]{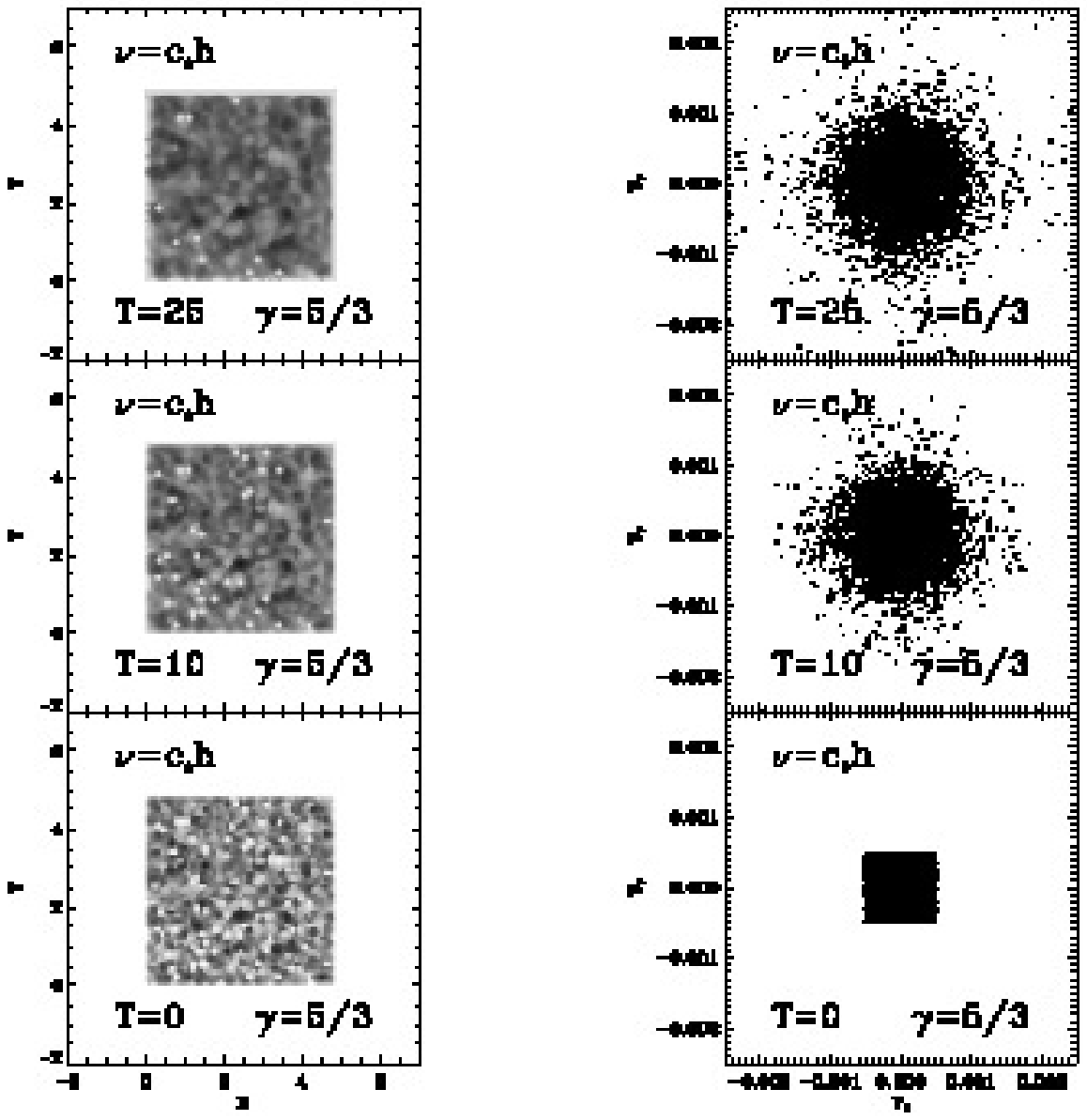}}
\caption{($X,Y$) plots as those of Fig. 2 for the physically viscous model $\nu = c_{s} h$.}
\end{figure}

\begin{figure}
\resizebox{\hsize}{!}{\includegraphics[clip=true]{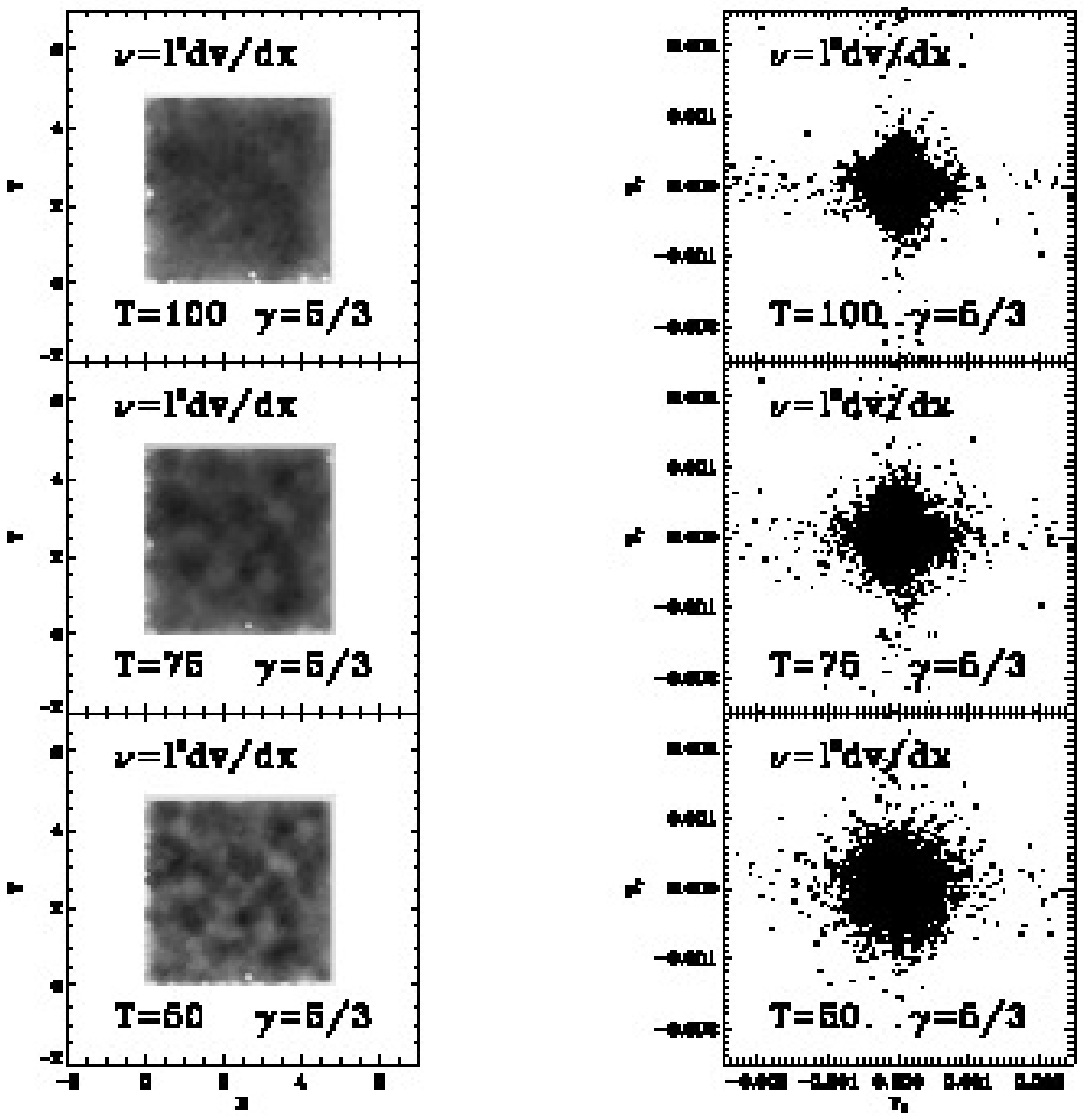}}
\resizebox{\hsize}{!}{\includegraphics[clip=true]{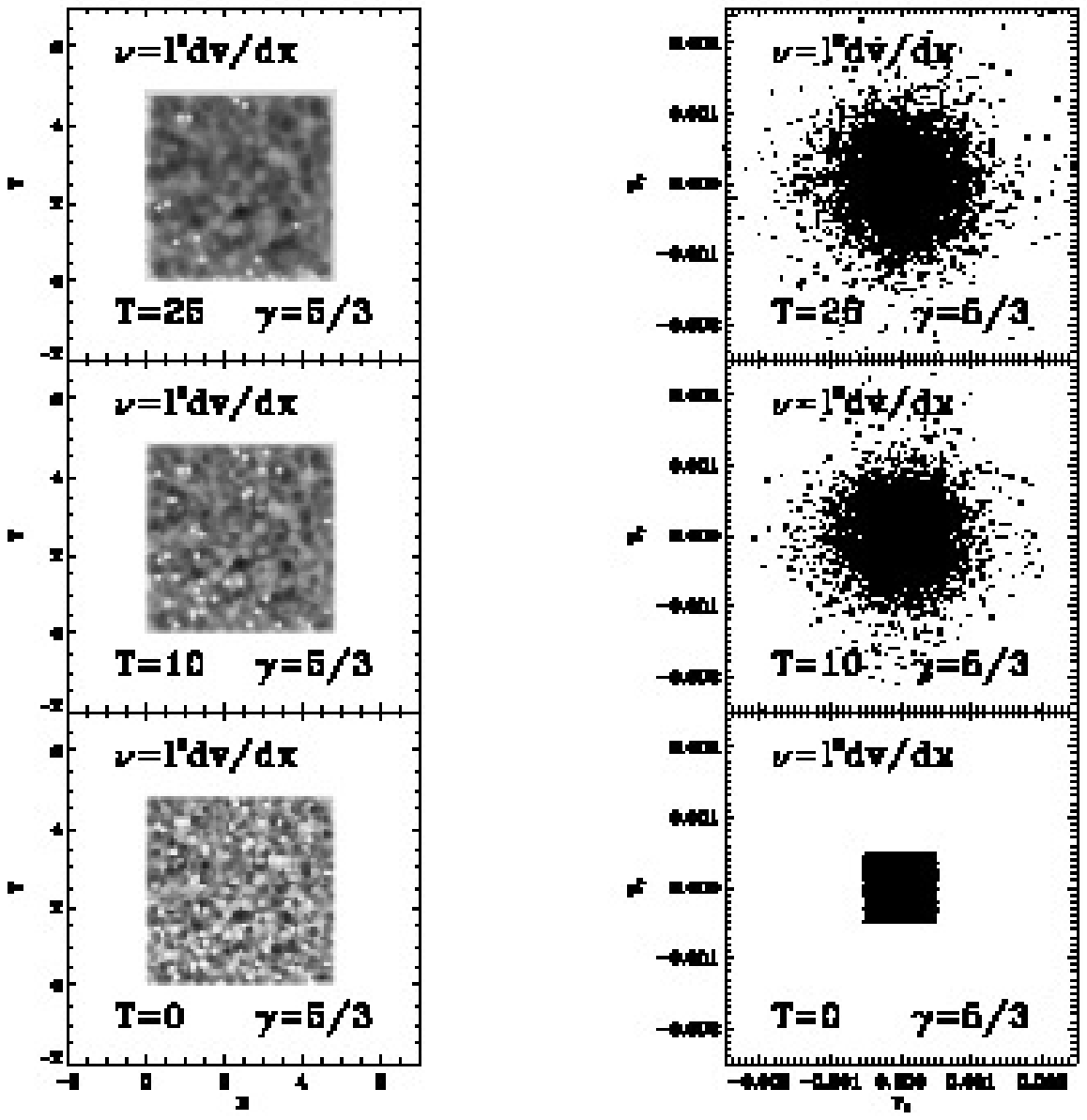}}
\caption{($X,Y$) plots as those of Fig. 2 for the physically viscous model $\nu \sim l^{2} \left|\frac{\partial v}{\partial x} \right|$.}
\end{figure}

\begin{figure}
\resizebox{\hsize}{!}{\includegraphics[clip=true]{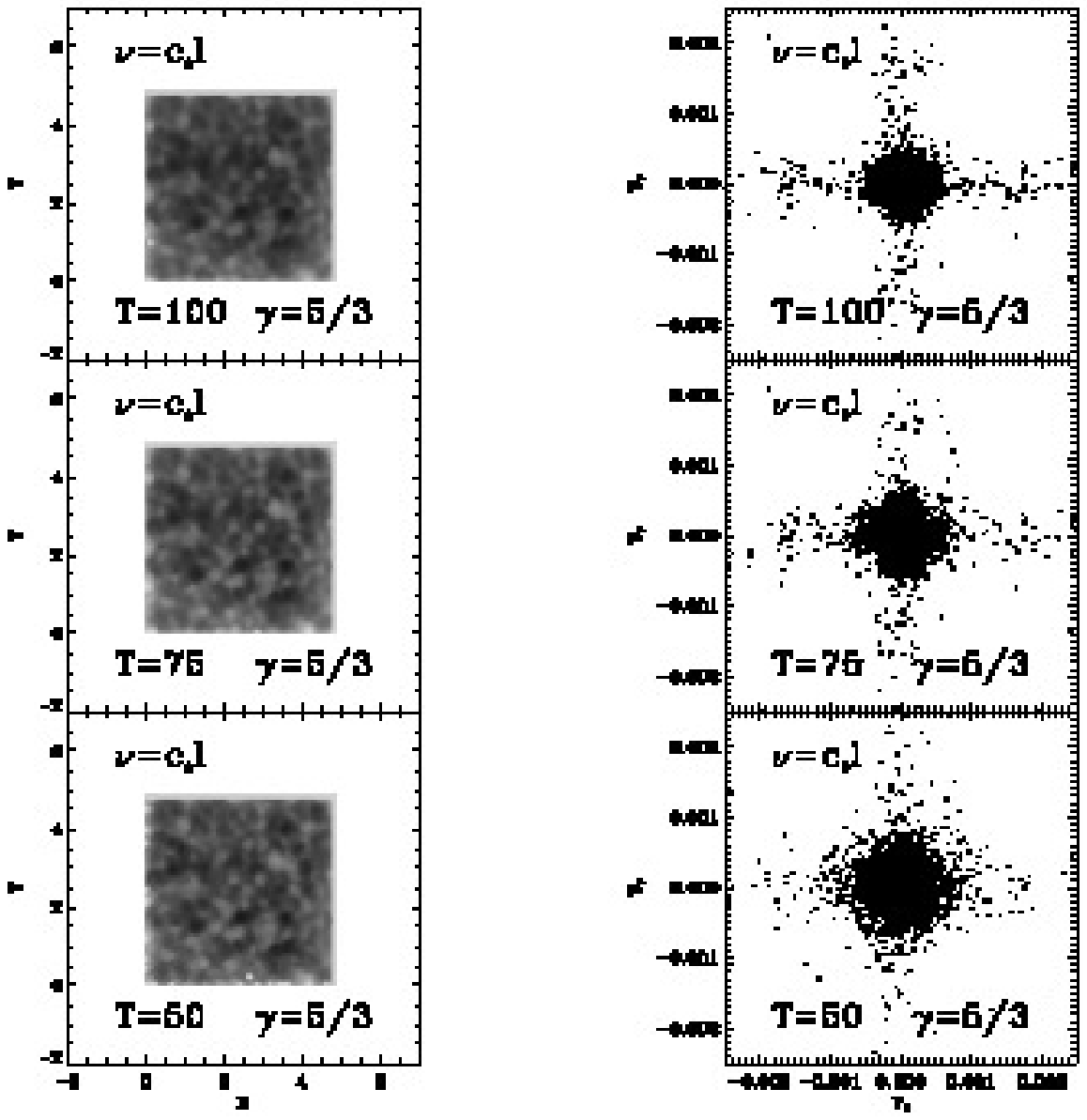}}
\resizebox{\hsize}{!}{\includegraphics[clip=true]{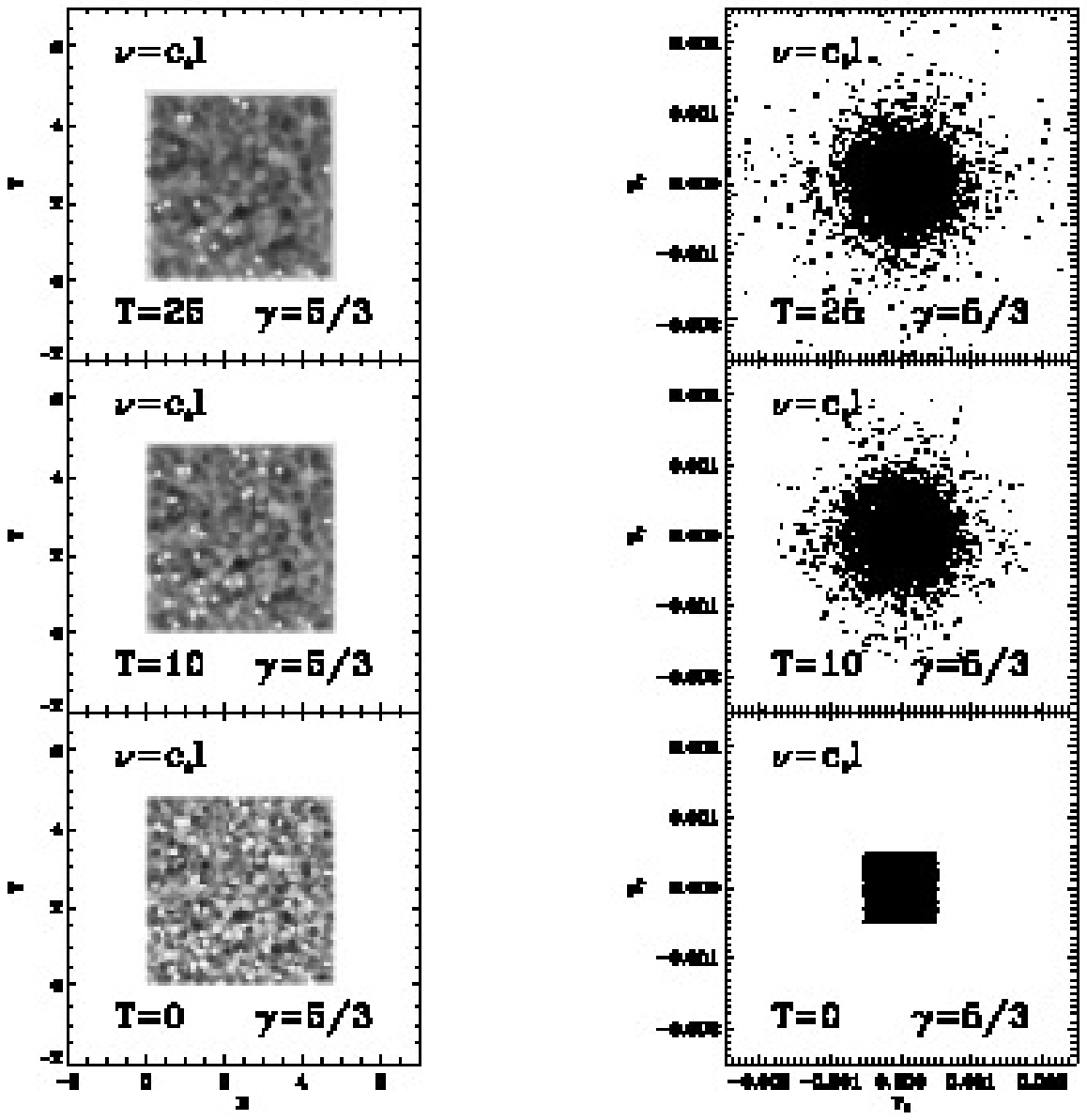}}
\caption{($X,Y$) plots as those of Fig. 2 for the physically viscous model $\nu = c_{s} l$.}
\end{figure}

\begin{figure}
\resizebox{\hsize}{!}{\includegraphics[clip=true]{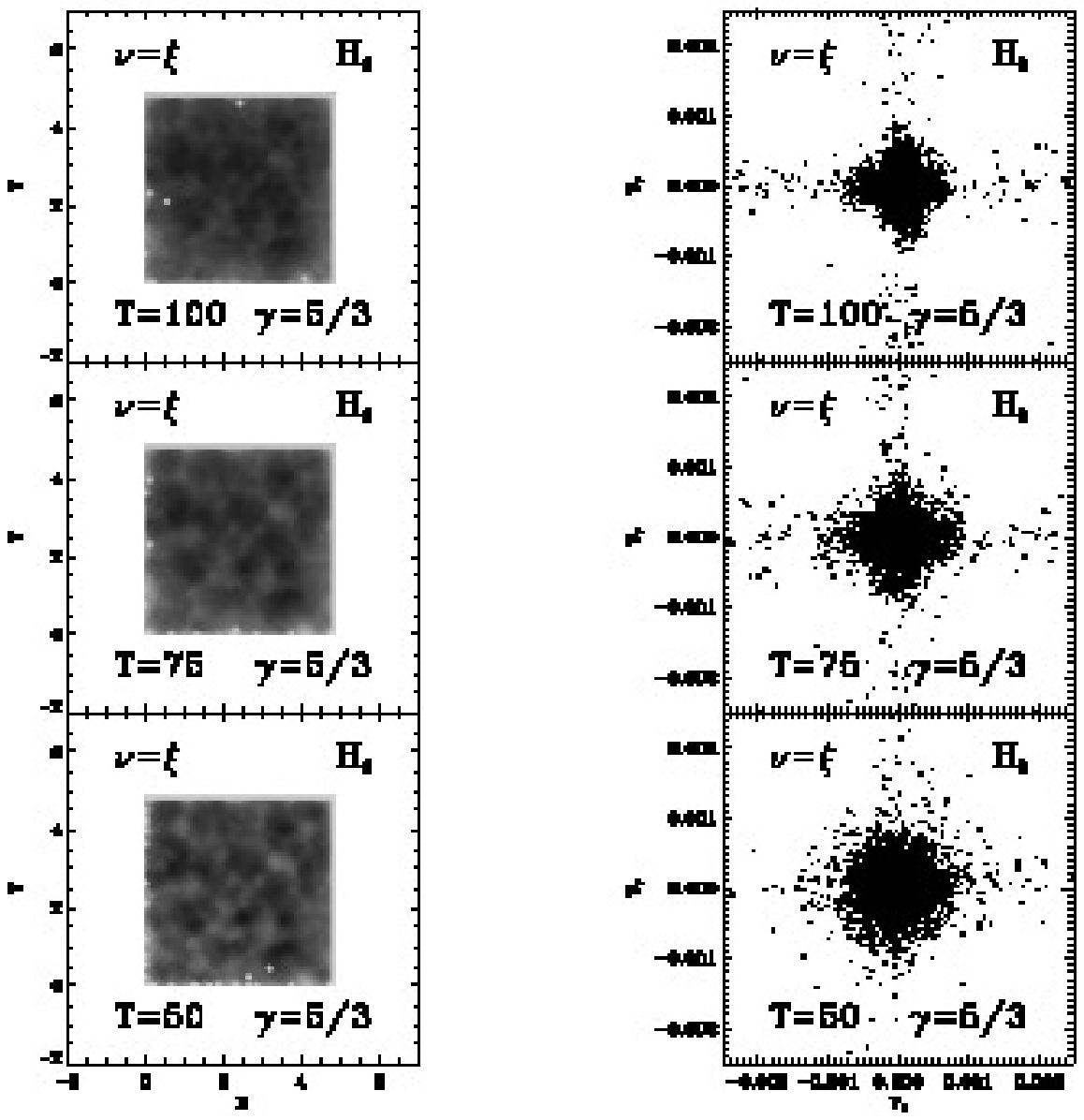}}
\resizebox{\hsize}{!}{\includegraphics[clip=true]{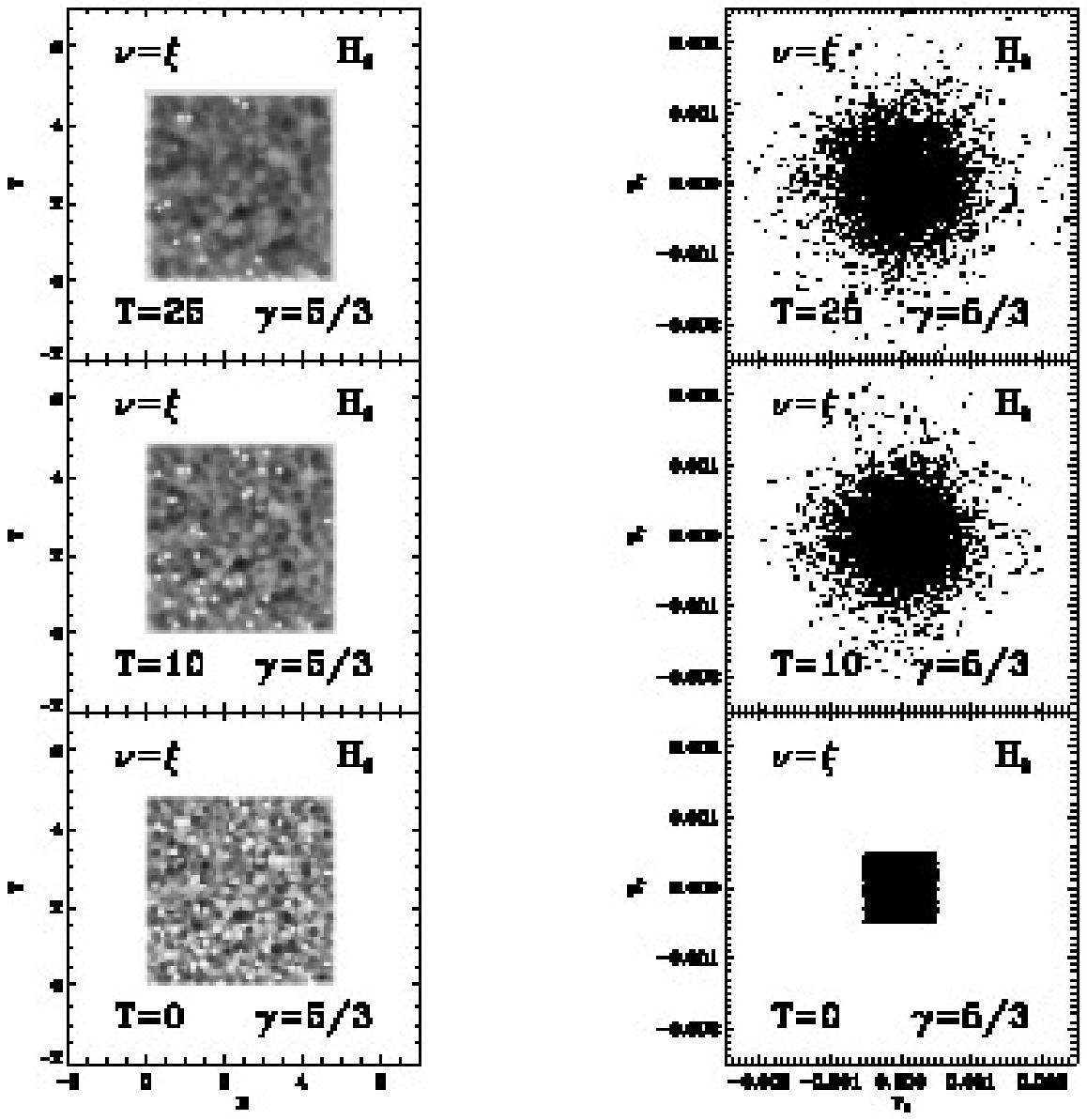}}
\caption{($X,Y$) plots as those of Fig. 2 for the physically viscous model $\nu = \xi$ for a $H_{2}$ gas. $\Sigma_{\circ} = 10^{-10}$ g cm$^{-2}$.}
\end{figure}

\begin{figure}
\resizebox{\hsize}{!}{\includegraphics[clip=true]{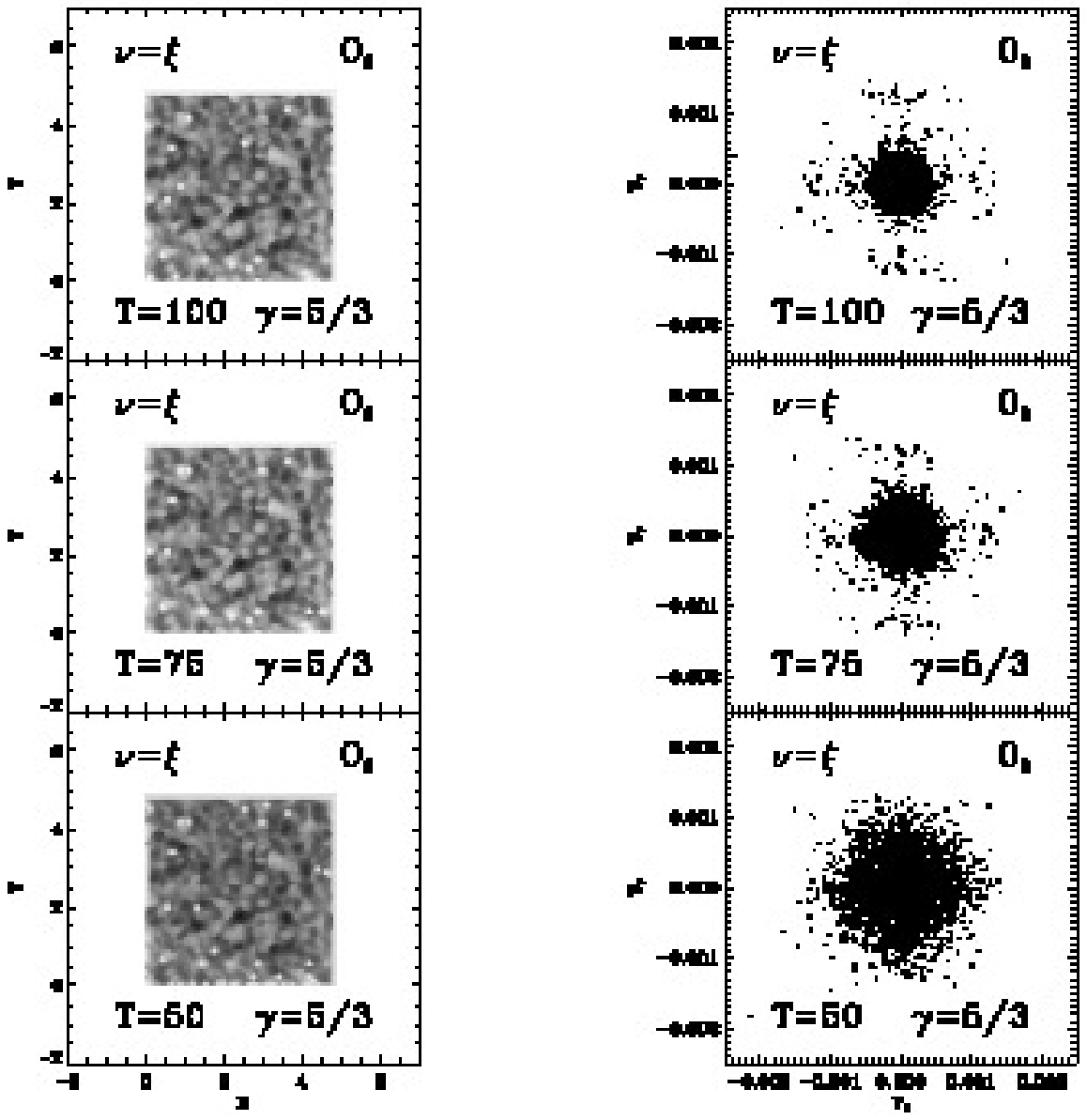}}
\resizebox{\hsize}{!}{\includegraphics[clip=true]{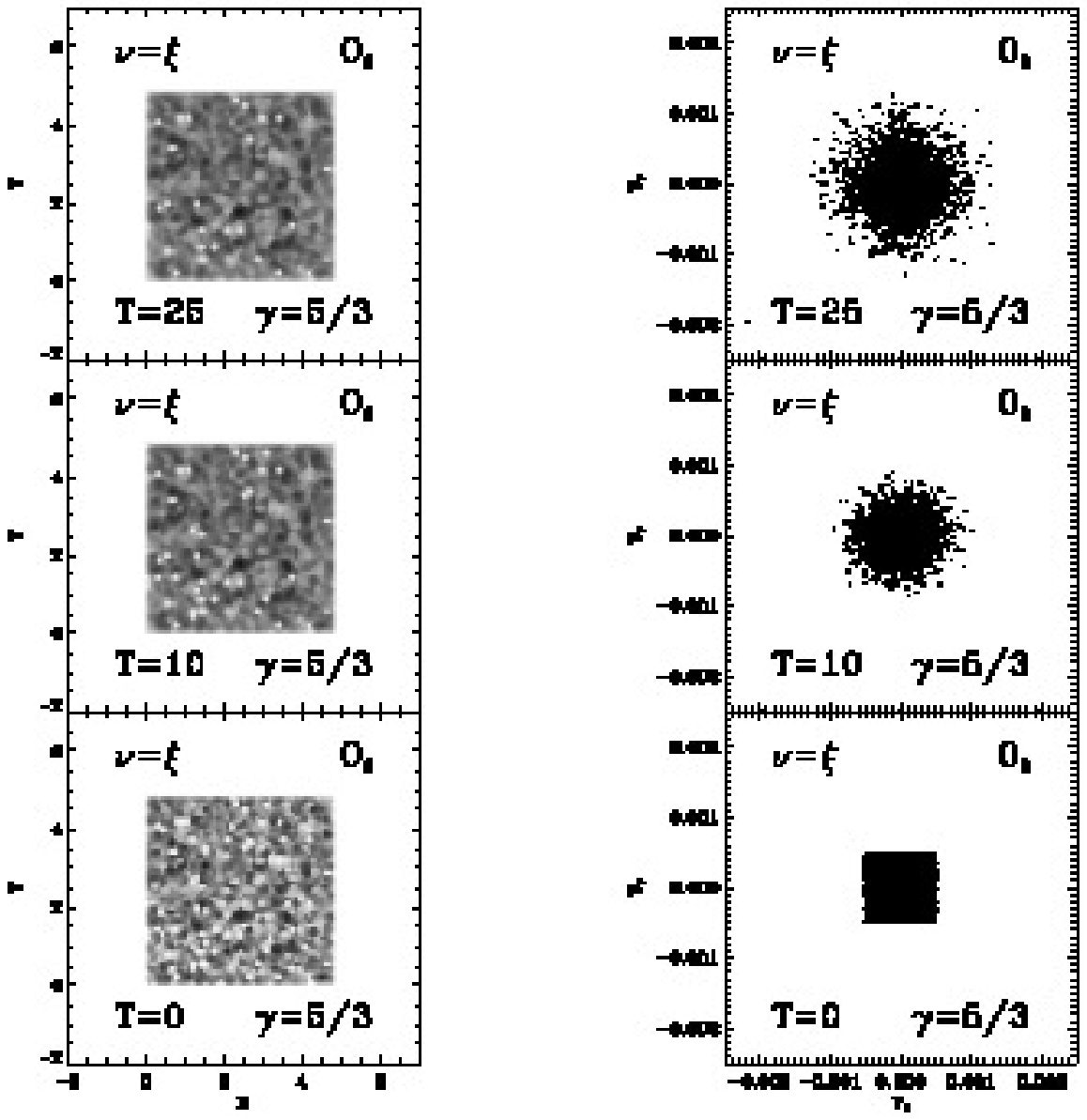}}
\caption{($X,Y$) plots as those of Fig. 2 for the physically viscous model $\nu = \xi$ for a $O_{2}$ gas. $\Sigma_{\circ} = 1.6 \cdot 10^{-9}$ g cm$^{-2}$.}
\end{figure}

\begin{figure}
\resizebox{\hsize}{!}{\includegraphics[clip=true]{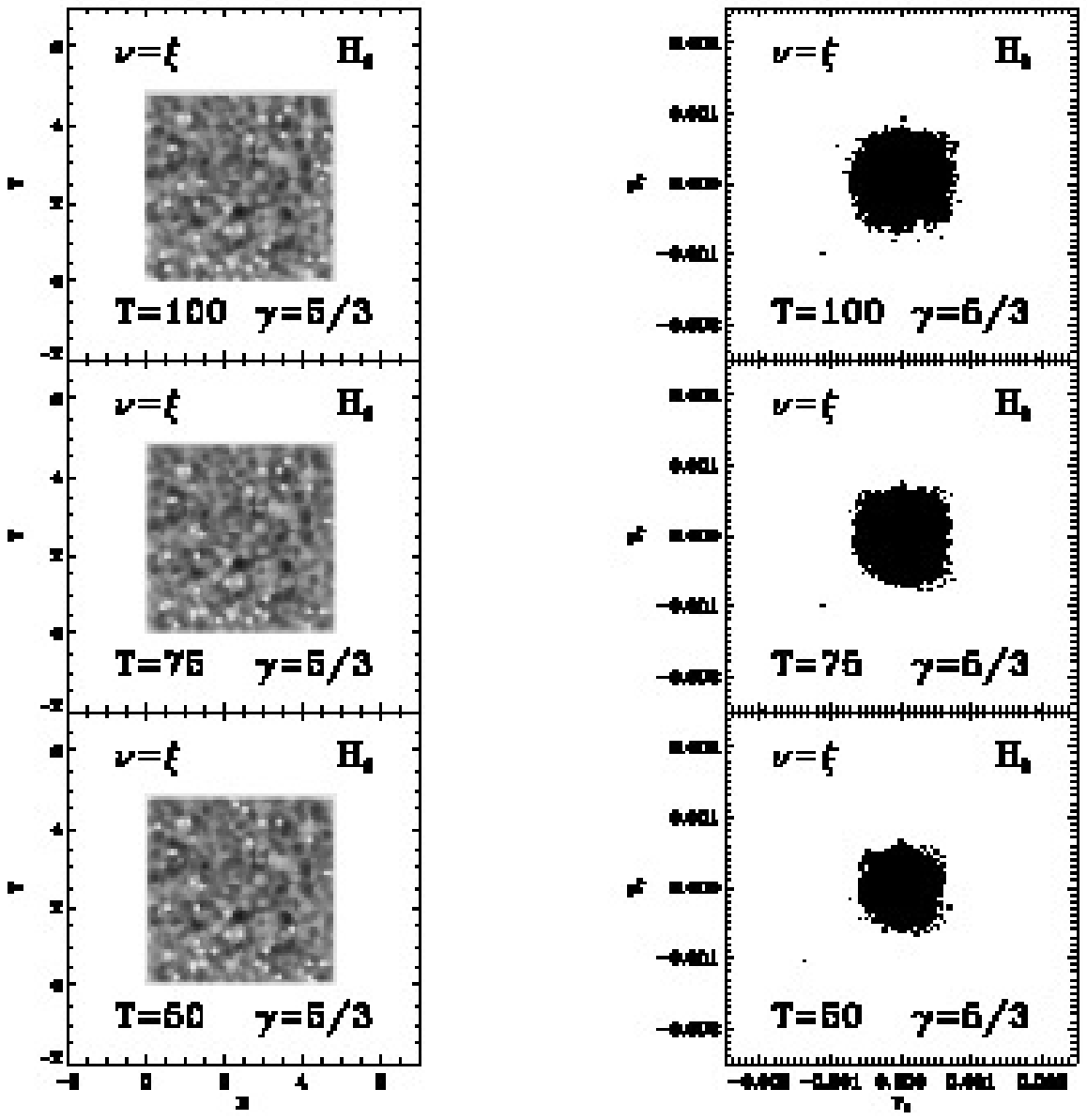}}
\resizebox{\hsize}{!}{\includegraphics[clip=true]{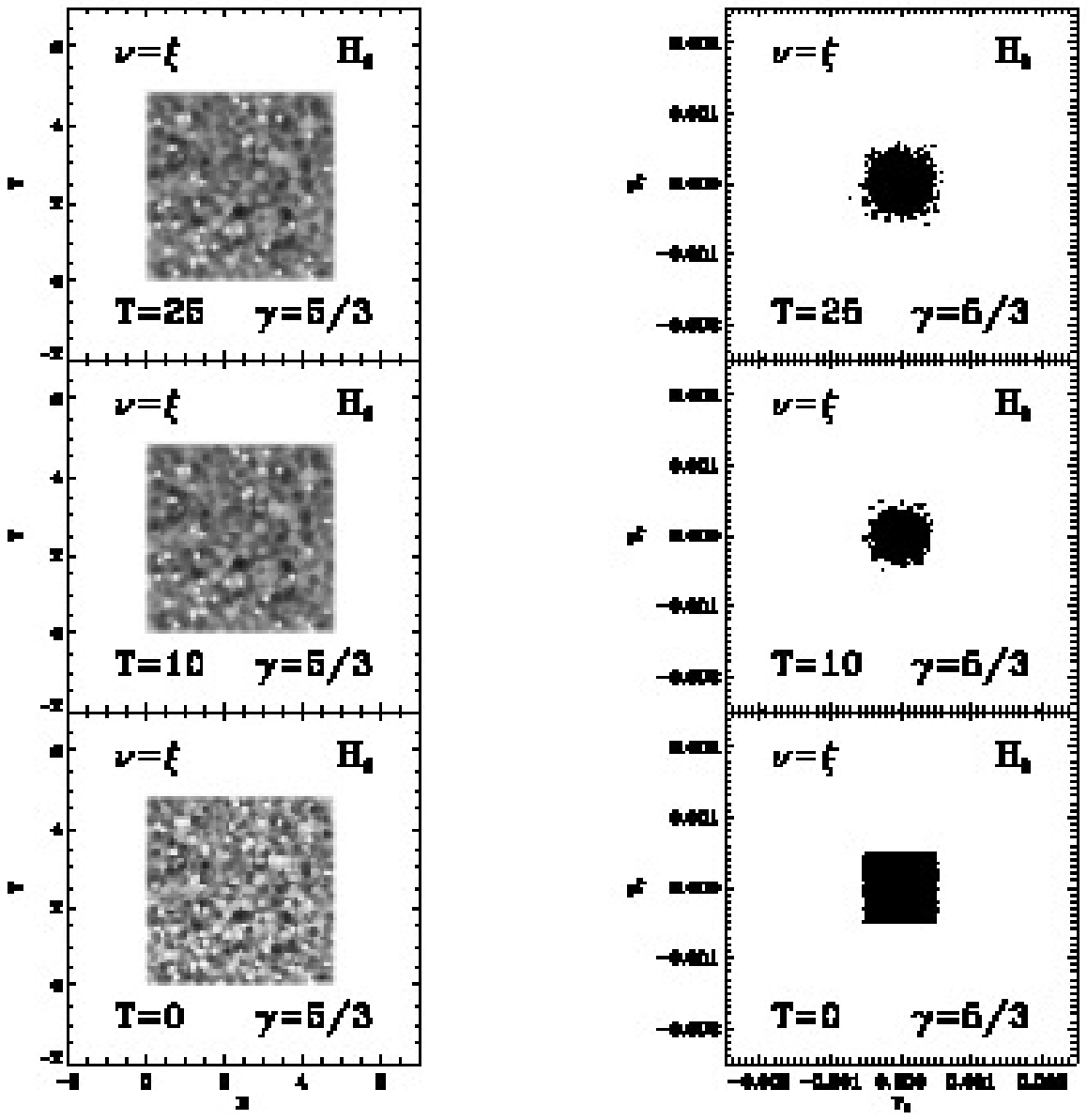}}
\caption{($X,Y$) plots as those of Fig. 2 for the physically viscous model $\nu = \xi$ for a $H_{2}$ gas. $\Sigma_{\circ} = 10^{-8}$ g cm$^{-2}$.}
\end{figure}

  2D phenomenology is somewhat more complex than 3D phenomenology, even though computationally more convenient \citep{a67}, being not derivable from simple dimensional arguments. 3D turbulence involves scales smaller than the trigger one and it is supposed to be hosted to a direct enstrophy cascade from large to small scales (from small to large wavenumbers), where the mean kinetic energy is transferred and mean enstrophy is conserved. Instead, 2D turbulence involves larger scales than the trigger one and it is supposed to be hosted to an inverse enstrophy cascade from small to large scales (from large to small wavenumbers). An inverse cascade of energy and a contemporary direct cascade of enstrophy in 2D is called a "dual" cascade \citep{a63}. Even though a physical viscosity is usually considered, as responsible of flow damping of the Navier-Stokes equations, the physical counterpart of artificial viscosity has been discussed in \citet{c37,c5,c3,c4}.

  Turbulence cascade theory substantially predicts that the kinetic energy is contained in turbulent eddies, while the cascade process for enstrophy is somewhat ambivalent in 2D, according to boundary and initial conditions. The larger the eddy, the larger its kinetic energy content, according to a power scaling law of eddy dimension, respecting the energy conservation law (kinetic + thermal) within the whole system. In 2D, the so called "dual cascade" process determines the formation of larger turbulent eddies up to the limit of the entire spatial domain.

  The numerical experiment here carried out for 2D Burger's turbulence studies the temporal evolution of the damping of a chaotic gas inside a $L = 5$ size 2D squared box, whose density and whose kinetic velocity are initially locally random, as an example of decaying turbulence. $v_{X}$ and $v_{Y}$ values are within the range $-5 \cdot 10^{-4}$ and $5 \cdot 10^{-4}$ at the beginning while $h = 0.05$. Hence $v_{X} \sim 0$ and $v_{Y} \sim 0$ statistically at time $T = 0$. Throughout the models the initial adimensional thermal energy per unit mass $\epsilon = 1.86 \cdot 10^{-7}$.

  The initial settings consists of an uniform thermal energy together with a statistical macroscopic homogeneous and isotropic spatial distributions on density and on velocity components $v_{X}$, $v_{Y}$. Being the macroscopic scale length much larger than the spatial resolving power, clear fluctuation exists in the homogeneity on the scale resolution length. This is enough to ignite the initial turbulent kinematics, with larger velocity, thermal energy and density fluctuations as larger are pressure forces from the beginning. Damping viscous effects always work since the beginning. Hence, as a decaying test, after the initial turbulent condition, where turbulence dominates, the final configuration is characterized by a fluid statistically at rest, where fluctuations in the velocity field, in the thermal energy and in the density are reduced after some time toward a more general uniformity.

  Figg. 2 to 6 show $XY$ distribution of density at various times $T$ for various physical kinematic turbulent viscosities $\nu$, as well as parallel velocity tomograms showing $v_{X}$ and $v_{Y}$ at the same time, from time $T = 0$ to $T \sim 100$. $64$ grey tones for the density maps are shown between the minimum and the maximum $\Sigma$ values for each plot. That is, each plot has its own minimum and maximum value for $\Sigma$. The normalization value for the density is $\Sigma_{\circ} = 10^{-10}$ g cm$^{-2}$ and densities are initially computed multiplying $\Sigma_{\circ}$ times a random number between $0$ and $1$. We distinguish two physical regimes. In the first one, from the beginning to $T \approx 25$, 2D turbulence dominates because small eddies grow up. Linear size of eddies, initially of the order of $h$, grows up to several $h$ values. The relevance of the artificial dissipation - $\nu \approx (0.1 - 0.4) c_{s} h$ - is evident in the physically non viscous case, because of the low spatial resolution adopted. From $T \sim 25$ onwards, clear differences both on the density spatial distribution and on the velocity component fluctuation appear, distinguishing the efficacy of the various adopted physical viscosity coefficients, since the physical viscosity better works increasing $c_{s}$, as well as enlarging the eddies if $\nu \propto l$ or $\propto l^{2}$. Plots clearly show that $\nu = c_{s} l$ is the most dissipative model in so far as $\Sigma_{\circ} = 10^{-10}$ g cm$^{-2}$. In this example, the physical damping relative to $\nu = \xi$ is comparable with that relative to the Prandtl's $\nu \sim l^{2} |\partial v\partial x|$.

  A reduction of the intrinsic damping by decreasing the spatial resolution length is possible, of course, but not practical. The reduction of the time step, together with a much larger computer memory needed involve very long computational time, even for 2D simulations. We adopted a low spatial resolution, since $h/L = 10^{-2}$. This involves that results inherent to all models are quite viscous. However, these results, even though conditioned by the intrinsic numerical damping effect, clearly show how $\xi$ correctly works, free of any dependence on $h$, or on any arbitrary parameter.

  To show the role of other chemical species, through their mean molecular weight, we also show in Fig. 7 results regarding the damping of 2D turbulence for a gas of molecular $O_{2}$ for $\Sigma_{\circ} = 1.6 \cdot 10^{-9}$ g cm$^{-2}$ at time $T = 0$. These results have to be compared with those for a gas of molecular $H_{2}$ with the same initial numerical density (Fig. 6), whose $\Sigma_{\circ} = 10^{-10}$ g cm$^{-2}$ at $T = 0$. Results for $O_{2}$ are influenced by the gas molecular cross section $\overline{\kappa}$, in so far as the numerical density $\rho/\overline{\mu} m_{H}$ is relevant in the $\xi$ calculation to get a $\nu$ comparable or higher than any artificial or any numerical dissipation. In this example, being the numerical densities for the molecular $O_{2}$ at the beginning the same as those of Fig. 6 for $H_{2}$, $\nu = \xi$ is so effective that viscous dissipation quickly dampens both any initial velocity fluctuation and flow progression toward any spatial homogeneity.

  An even stronger viscous flow damping is shown in Fig. 8 for a denser gas of molecular $H_{2}$, whose $\Sigma_{\circ} = 10^{-8}$ g cm$^{-2}$ at $T = 0$. Viscous molecular damping, where $\nu = \xi$, appears more effective for higher density flow modelling. As an example, in Fig. 8 viscous dissipation is so relevant that any local flow motion is strongly suppressed since the beginning, also because of a more relevant heat thermal conductivity suppressing local thermal differences caused by the viscous heating. At the same time, the viscous heating is so strong that the consequent increase of pressure forces time by time activates again the velocity fluctuations in the flow after $T \sim 50$. Hence, a different evolution of velocity fluctuation is developed, depending on the initial conditions for $v$, $\epsilon$, $\nu$, $\Sigma$. The study of the 2D viscous damping of the 2D kinematics of a so dense fluid does not need to concern us.

\section{3D accretion disc in a close binary}

\begin{figure*}
\resizebox{\hsize}{!}{\includegraphics[clip=true]{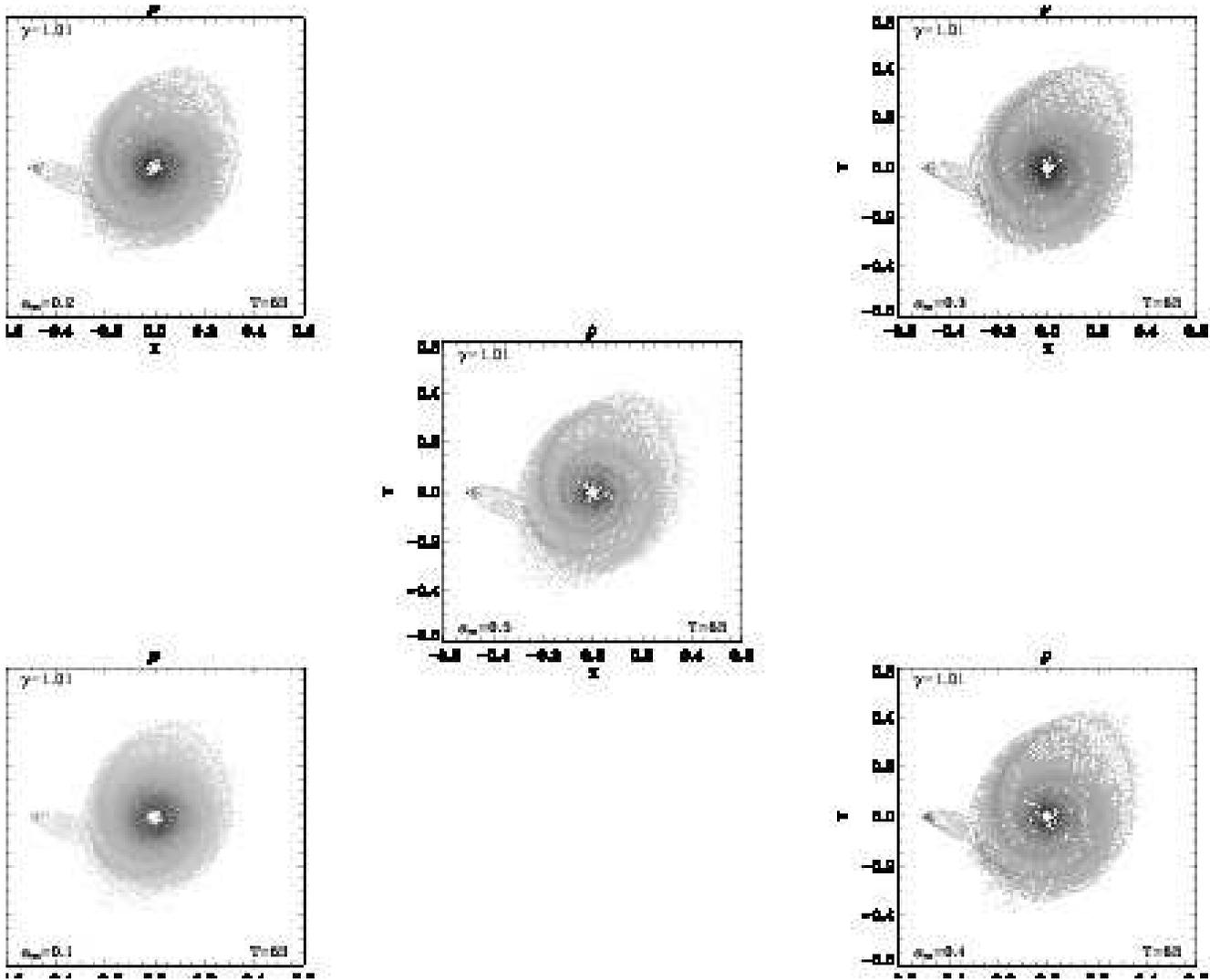}}
\caption{($X,Y$) plots for $64$ greytones of density map for 3D physically viscous accretion disc in the $\nu = \alpha_{SS} c_{s} H$ Shakura and Sunyaev formulation for $\gamma = 1.01$. Time $T$ and  $\alpha_{SS}$ are also shown.}
\end{figure*}

\begin{figure*}
\resizebox{\hsize}{!}{\includegraphics[clip=true]{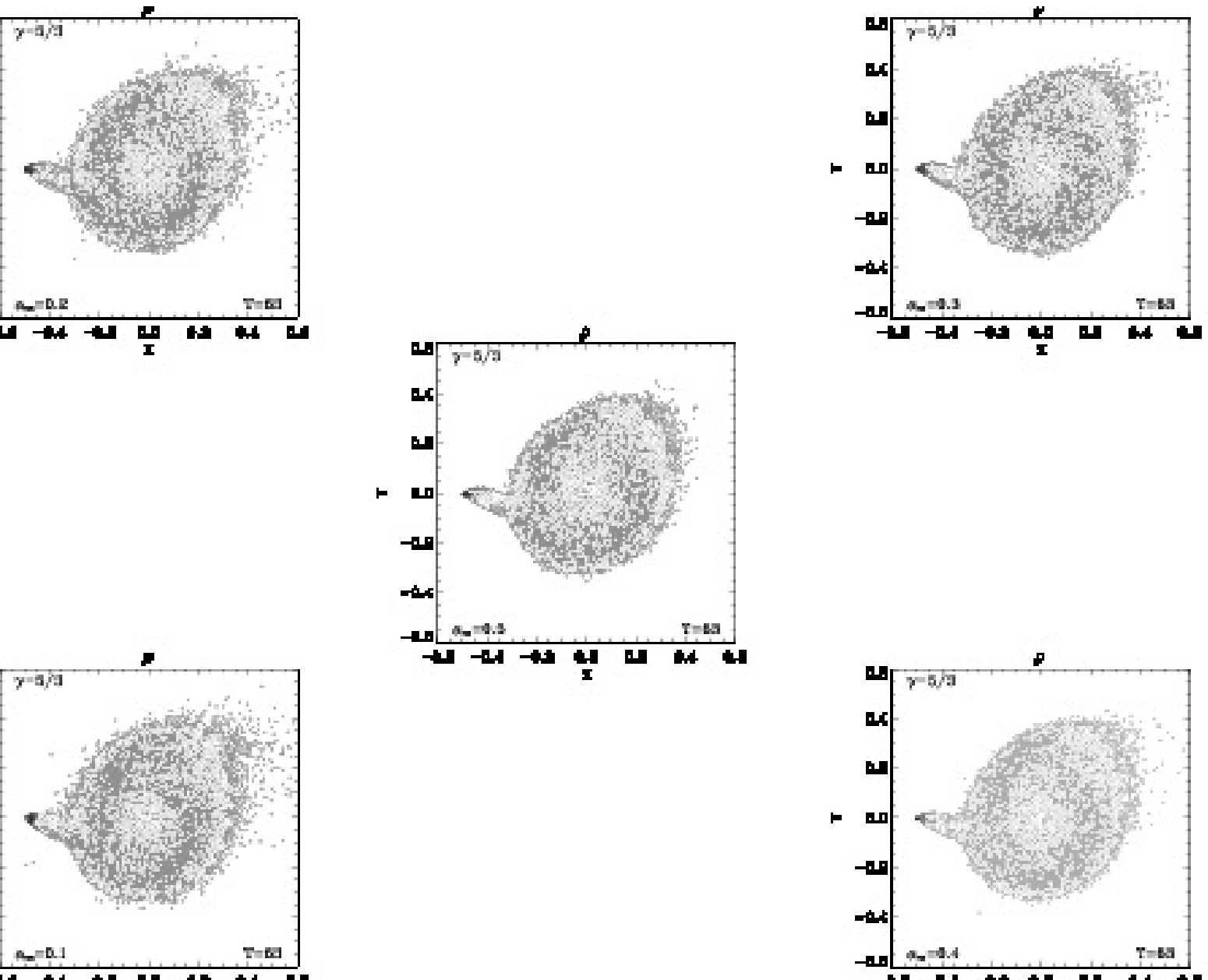}}
\caption{($X,Y$) plots for $64$ greytones of density map for 3D physically viscous accretion disc in the $\nu = \alpha_{SS} c_{s} H$ Shakura and Sunyaev formulation for $\gamma = 5/3$. Time $T$ and  $\alpha_{SS}$ are also shown.}
\end{figure*}

\begin{figure*}
\resizebox{\hsize}{!}{\includegraphics[clip=true]{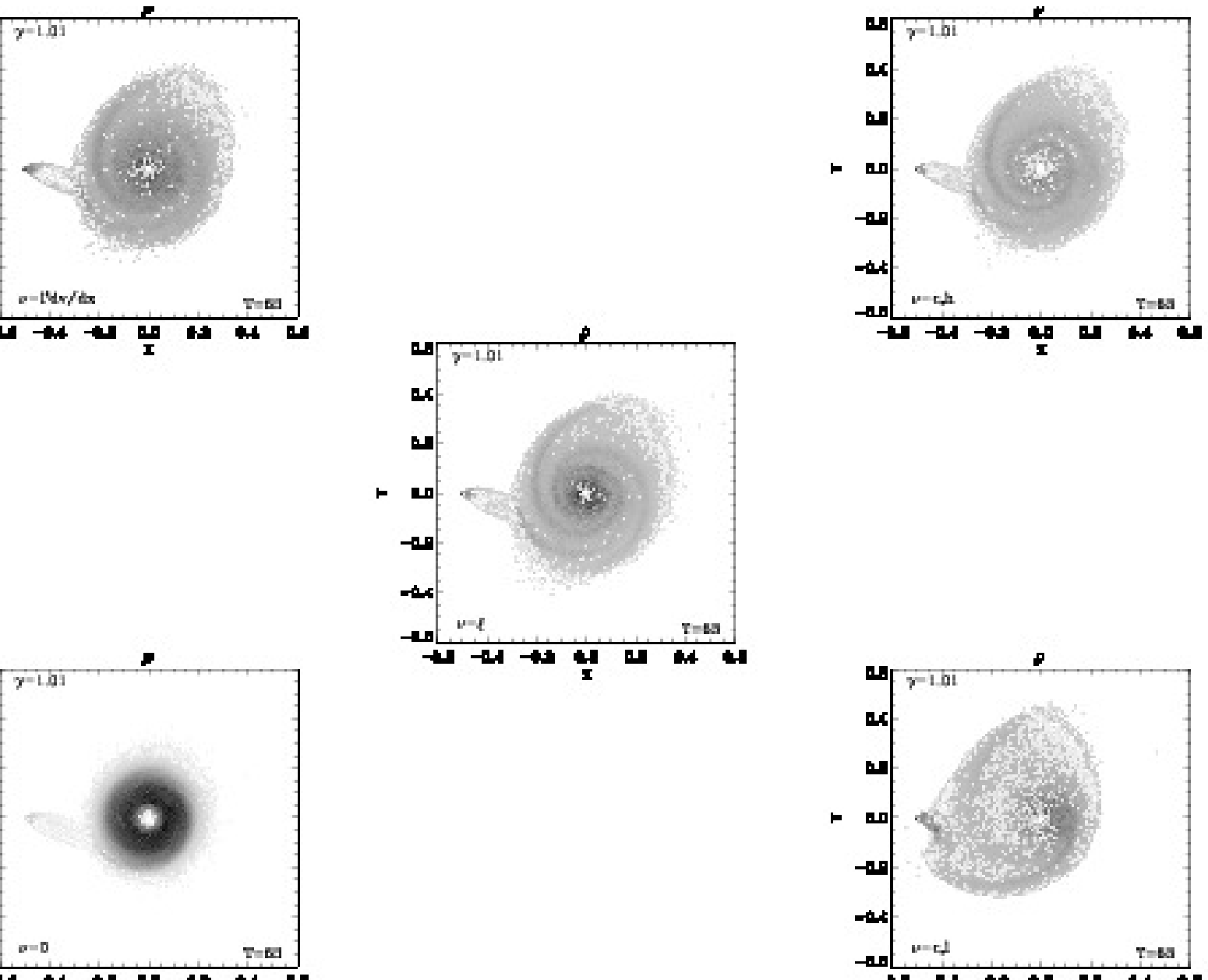}}
\caption{($X,Y$) plots for $64$ greytones of density map for 3D physically viscous accretion disc for $\gamma = 1.01$. Time $T$ as well as the used $\nu$ formulation are also shown.}
\end{figure*}

\begin{figure*}
\resizebox{\hsize}{!}{\includegraphics[clip=true]{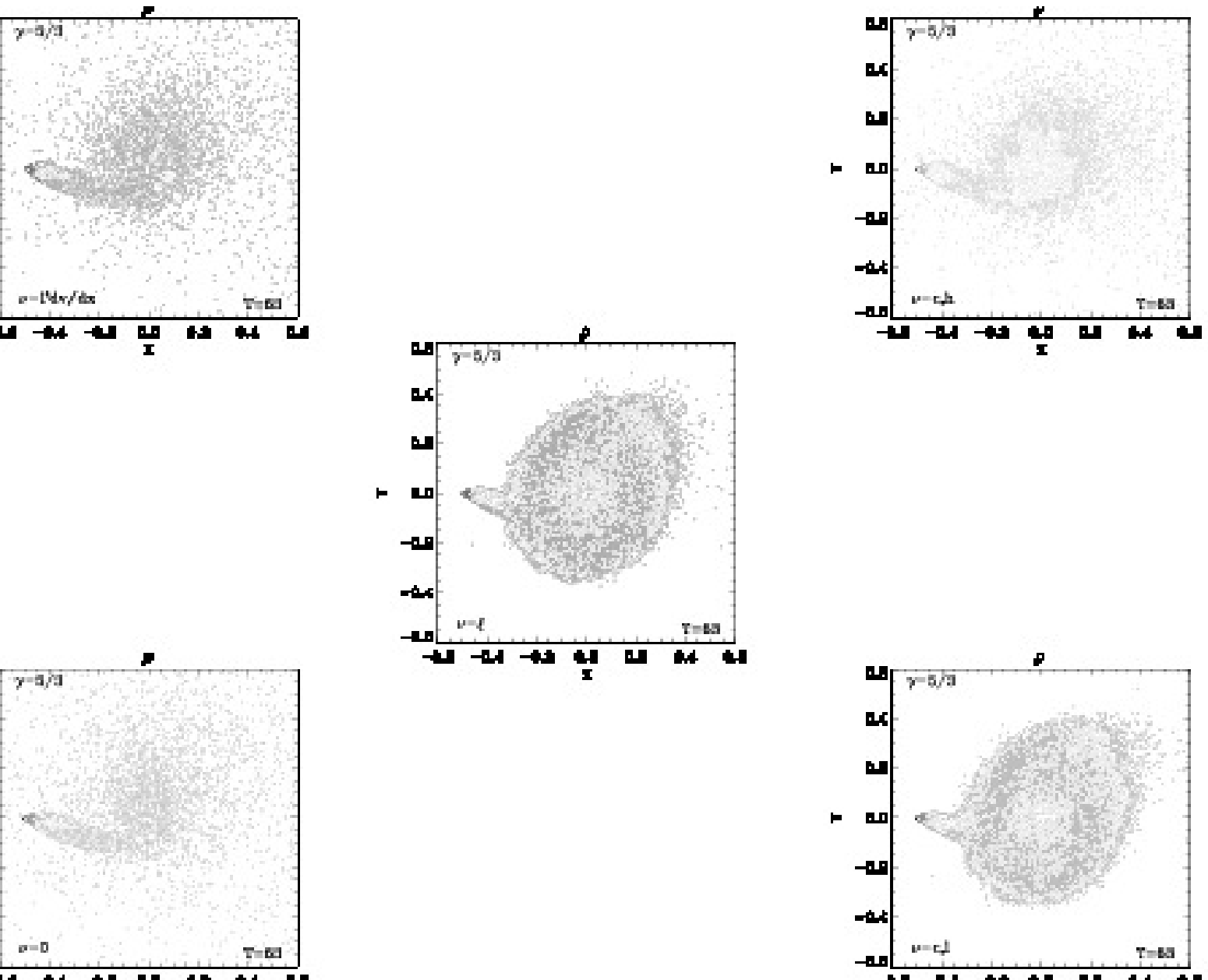}}
\caption{($X,Y$) plots for $64$ greytones of density map for 3D physically viscous accretion disc for $\gamma = 5/3$. Time $T$ as well as the used $\nu$ formulation are also shown.}
\end{figure*}

\begin{figure*}
\resizebox{\hsize}{!}{\includegraphics[clip=true]{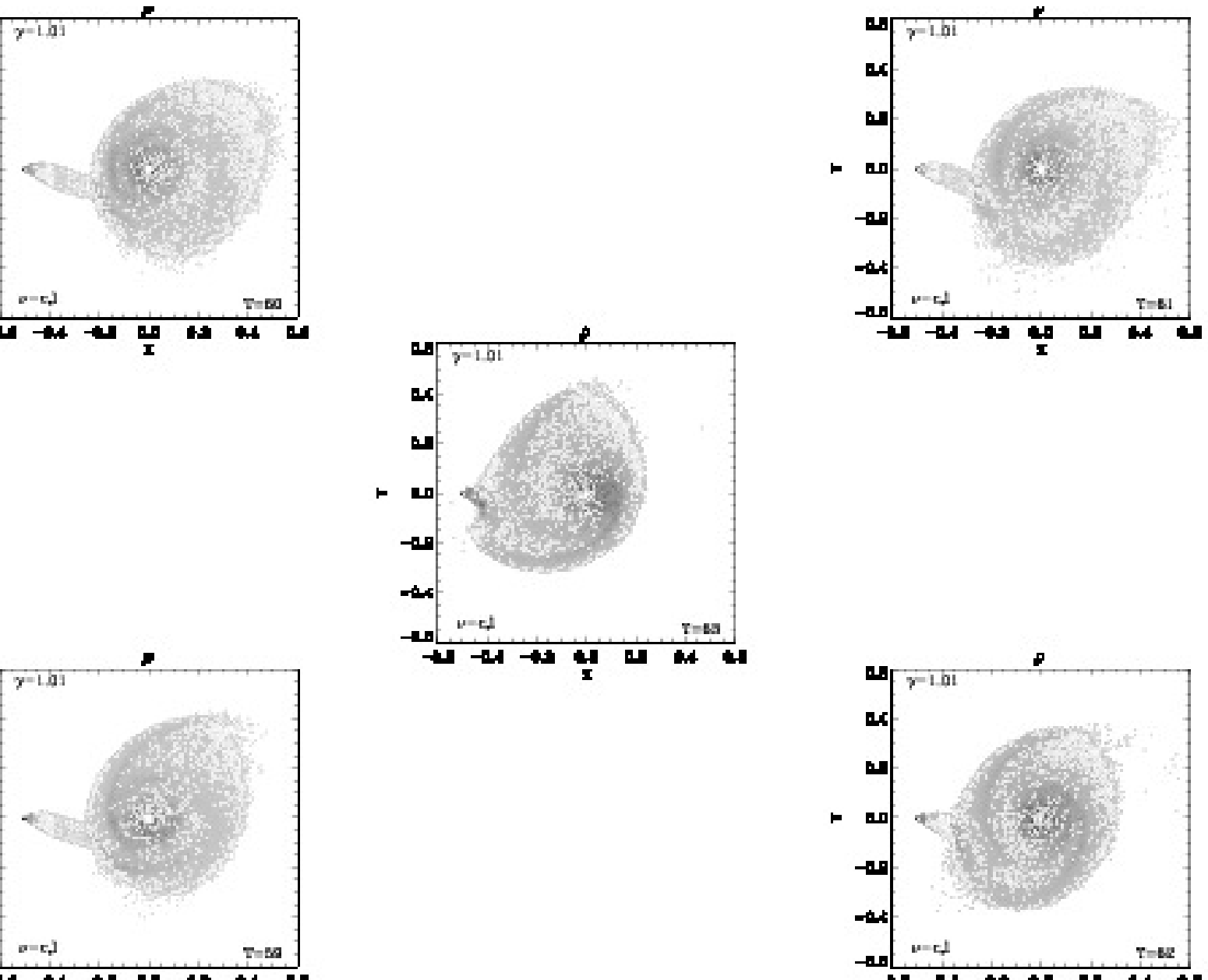}}
\caption{($X,Y$) plots for $64$ greytones of density map for $5$ instants of the 3D the physically viscous accretion disc for $\gamma = 1.01$, $\nu = c_{s} l$. Time $T$ is also shown.}
\end{figure*}

  We compare results for 3D stationary disc structures both in high compressibility ($\gamma = 1.01$) and in low compressibility ($\gamma = 5/3$) with the aim of getting a physically well-bound accretion disc around the compact primary star in a LMCB. However, these comparisons are thought to show which physically viscous disc structure, in the Shakura and Sunyaev formulation (eq. 22), in the Prandtl formulation (eqs. 16, 17), and in the more simple formulations (eqs. 32 and 33), better compare with that relative to $\nu = \xi$. The adopted spatial resolution length is $h = 5 \cdot 10^{-3}$ throughout the simulations, being the mutual separation of the two stars normalized to $1$.

  The characteristics of the binary system are determined by the masses of the primary compact white dwarf and of its companion star and their separation. We chose to model a system in which the mass $M_{1}$ of the primary compact star and the mass $M_{2}$ of the secondary subgiant star are both equal to $1 M_{\odot}$ and their mutual separation is $d_{12} = 10^{6} \ Km$. The injection gas velocity at L1 is fixed to $v_{inj} \simeq 130$ Km s$^{-1}$ while the injection gas temperature at L1 is fixed to $T_{\circ} = 10^{4} \ K$, taking into account, as a first approximation, the radiative heating of the secondary surface due to radiation coming from the disc. Supersonic kinematic conditions at L1 are discussed in \citet{c45,c46,c47,c59}, especially when active phases of CB's are considered. The reference frame is centred on the primary compact star, whose rotational period, normalized to $2 \pi$, coincides with the orbital period of the binary system, being the velocities normalized to $v_{\circ} = [G (M_{1} + M_{2}/d_{12})]^{1/2}$. Results of this paper are to be considered a useful test to check whether disc structures (viscous and non) show the expected behaviour.

  We simulated the physical conditions at the inner and at the outer edges as follows:

a) inner edge: \\
the free inflow condition is realized by zeroing gas flow inside a sphere of radius $10^{-2}$, centred on the primary compact star. Although disc structure and dynamics are altered near the inner edge, these alterations are relatively small because they are balanced by a high gas concentration close to the inner edge in supersonic injection models.

b) outer edge: \\
gas flow from L1 towards the interior of the primary Roche Lobe is simulated by constant gas pressure, density, and thermal energy per unit mass, as well as a constant velocity in a small conic region having L1 as a vertex and an aperture of $\sim 57^{\circ}$. The radial length of this small volume is $\sim 10 h$. The initial injection particle velocity is radial with respect to L1. Local density at the inner Lagrangian point L1: $\rho_{L1} = 10^{-14}$ g cm$^{-3}$. This order of magnitude is explained as follows: from the flux conservation, $\rho v S =$ const. Hence, $\rho_{1} v_{1} S_{1} = \rho_{2} v_{2} S_{2}$ from the two Roche lobe sides at L1. If the two stars have comparable masses, $\rho_{1} v_{1} \simeq \rho_{2} v_{2}$. If, at L1, from the secondary subgiant star side, $\rho_{2} \approx 10^{-11}$ g cm$^{-3}$ and $v_{2} \approx 10^{-1}$ Km s$^{-1}$ as typical photospheric values for a subgiant, then $\rho_{1} \approx 10^{-14}$ g cm$^{-3}$, being $v_{1} \approx 10^{2}$ Km s$^{-1}$.

  Supersonic mass transfer conditions from L1 were previously adopted in \citet{c45,c46}, where disc instabilities, responsible for disc active phases of CB are discussed in the light of local thermodynamics. Whenever a relevant discrepancy exists in the mass density across the inner Lagrangian point L1 between the two stellar Roche lobes, a supersonic mass transfer occurs as a consequence of the momentum flux conservation. The same result can also be obtained \citep{c49} by considering either the restricted problem of three bodies in terms of the Jacobi constant or the Bernoulli's theorem. Although the stellar gas is surely neutral for $T = 10^{4} \ K$ at L1, we consider a chemical composition of pure ionized hydrogen for the sake of simplicity for a simple calculation of $\overline{\kappa}$ in the disc bulk.

  Figg. 9 and 10 show five XY plots of the physically viscous discs for $\gamma = 1.01$ and for $\gamma = 5/3$, respectively, in the Shakura and Sunyaev prescription, where $\alpha_{SS}$ ranges from $0.1$ to $0.5$ in steps of $0.1$. Each density map has its own minimum and maximum densities, scaled in $64$ greytones. These $\alpha_{SS}$ values are in accordance with the typical and with the maximum compatible with both astrophysical observations \citep{c52} and with numerical experiments \citep{c53} for these kind of astrophysical objects. As it is evident, this formulation always works quite well, complicating the determination of the right value for the $\alpha_{SS}$ arbitrary parameter up to the point that viscous disc structures are practically indistinguishable from each other for $\gamma = 5/3$.

  Figg. 11 and 12 also compare density maps for $\gamma = 1.01$ and for $\gamma = 5/3$, respectively, both for the physically inviscid and for the other four physically viscous formulations for $\nu$.

  In the non viscous model, for $\gamma = 1.01$, Fig. 11 shows a compact disc structure, where a high density central torus is formed in the disc bulk. In this case, the intrinsic-artificial viscosity alone favours the radial mass viscous transport, as well as the conversion of mechanical energy into heat. A glimmer of spiral shape structures at the disc outer edge exists, although not well developed because of the limited radial extension of the disc together with its substantial circular symmetry. The tight link between the elliptical extension of eccentric disc and its asymmetry to the presence of spirals coming from its outer edge has been discussed by \citet{a1,a2,a3,a4,a5}, especially in relation to the "tidal truncation radius" \citep{a6,a7,a8,a10}.

  Notice in Fig. 11 for $\gamma = 1.01$ that, for a gas of pure ionized hydrogen, the viscous disc structure whose $\nu = \xi$ tightly compares with that whose $\nu = \alpha_{SS} c_{s} H$, where $\alpha_{SS} = 0.3 - 0.5$ (in Fig. 9). In the same figure 11, physically viscous discs, whose $\nu \sim l^{2} |\partial v/\partial x|$, $\nu = c_{s} h$ and $\nu = \xi$, are also elliptically extended, with a clear evidence of spiral structures coming from the eccentric disc outer edge from its more lengthened zone and from the disc side where the flow stream coming from L1 collides to the disc outer edge. For $\nu \sim l^{2} |\partial v/\partial x|$, the existence both of a disc thickness and of collisional spatial derivatives limits $l$, and consequently $\nu$, to local values not so large as those relative to the 2D shockless radial viscous transport for an annulus ring in \S 5.1.

  In 3D, despite the better limitation for $l$, the tendency toward a larger dissipation, as previously seen in \S 5.2 for the damping of Burger's turbulence for 2D structures (or for flattened 3D structures), gives the $\nu = c_{s} l$ disc structure a quite viscous behaviour. In the highly compressible and highly viscous case, not only the disc radial extension enlarges, but at the same time the entire structure better shows a retrograde precession in a waving of its outer edge, according to the reference frame of Fig. 13. This retrograde precession is originated by the gravitational tidal forces and supported by a significant Coriolis acceleration caused by the strong shockless radial viscous transport in a spiral-shape kinematics. Notice how in the momentum equation, the Coriolis acceleration is the only term depending on the radial component of $\bmath{v}$. The role of the Coriolis acceleration in the exchange of fluid blobs in disc structures is well described in the \citet{a69} textbook. The high viscosity and the high compressibility work together producing a gas structure comparable with that of a soft rubbish elastic membrane. Being the fluid highly deformable, but at the same time structurally bound, it is highly characterized by uniform collective motions and by an efficient conductive thermal transport, where any perturbation, even though strongly attenuated, propagates and involves as many fluid parts as possible. As a consequence, disc outer edge distortions are much more stressed. The disc distortion and the consequent lengthening of its outer edge better enhances the effect of tidal forces on the entire disc fluid dynamics. Results showing a retrograde clockwise precession in the viscous disc outer edge in a longitudinal oscillation distorting the entire disc profile were obtained by \citet{a20,a21,a33}, also in high artificial viscosity conditions, for LMCBs. Hence, a high viscosity together with a high compressibility gas are the essential conditions to get accretion discs where disc outer edge distortion and its tidal precession are evident in opposition to the standard disc models where such dynamics is absent. For $\nu = c_{s} l$ disc model, what is anomalous is the short period of waving of its distorted outer edge in a LMCB. It looks like comparable with the orbital period. Whenever in a LMCB a retrograde precession of the disc outer edge exists, it is at least $10$ times longer than the orbital period \citep{a20,a21,a33}, especially if tidal forces activate both the disc warping and its outer edge precession \citep{a34,a35,a36,a37}. This anomaly is however simply explained considering that for $\nu = c_{s} l$ the viscous time scale $\sim r^{2}/\nu$ is much shorter with respect to the viscous time scales of other less viscous disc modelling.

  Disc phenomenology is different in the low compressibility $\gamma = 5/3$ modelling, as shown by the comparison between Figg. 10 and 12. In the non viscous case, pressure forces as well as mass outflow from the disc outer edge are so relevant, that the entire disc structure is sparse and not well bound within the primary's gravitational potential well of a LMCB, a consolidated result since works of \citet{c48,a38}. The modest physical viscosity from the Prandtl's and from $\nu = c_{s} h$ does not allow any binding of the flow into the primary's gravitational potential well. Only a tiny gas concentration toward the inner regions of the potential well is visible. Instead, the two disc models for $\nu = \xi$ and for $\nu = c_{s} l$ look like working quite well, being also in a good comparison with the low compressibility viscous discs in the Shakura and Sunyaev prescription. In these two highly viscous collisional disc models, the larger physical viscosity damping, also evidenced in the 2D viscous damping of Burger's turbulent test, as well as an efficient thermal conduction smoothing out thermal and pressure gradients, determines an effective disc binding within the gravitational potential well. Moreover, the relevant high pressure forces for $\gamma = 5/3$ prevent any amplified disc oscillation of its outer edge as previously evidenced in Fig. 11 for $\gamma = 1.01$ and $\nu = c_{s} l$.

\section{Concluding remarks}

  In this paper we formulate a physical general expression for the kinematic viscosity coefficient $\nu$, able to determine a correct physical viscous flows in the nonlinear Navier-Stokes fluid dynamics. In this formulation, we search for an expression, free of any arbitrary numerical parameter, paying attention to the correlation between the microscopic molecular-atomic cross section to macroscopic characteristic lengths (mixing length) related to the linear eddy dimension. At the same time, a reformulation for $\nu$ in more strictly physical terms also involves a reformulation for the thermal conductivity coefficient $c$, being $\nu$ and $c$ simply correlated for a dilute gas.

  Current adopted kinematic viscosity coefficients are normally written as a pure mathematical coefficients, without any correlations to the molecular cross sections. Sometimes, as for example in the Shakura and Sunyaev formulation, limited only to disc geometric structures of the flow, an unknown arbitrary parameter appears. 

  Current $\nu$ formulations are not used whatever is the viscous physical problem. Either $\nu$ is specific for shockless viscous transport phenomena, or it is specific for chaotic turbulent flows. Of course in both cases, the conversion of mechanical energy into heat, together with the braking kinematics, yield laminar flows after some time without any external force acting on the flow.

  In this paper we also pay attention to the role of the intrinsic flow damping. It could be either explicit, as an artificial viscosity term, or it could be numerical, or both, especially for finite difference codes, because of second order terms by the numerical conversion of spatial derivatives coming from the Taylor series expansion. This non physical damping is necessary to handle collisional shocks, preventing any distortion in the shock fronts. In both situations, the adopted spatial resolution $h$ has a direct or indirect role.

  For practical reasons, we worked in a low spatial resolution, adopting $h/L = 10^{-3} - 10^{-2}$ throughout. This means that our results are affected by a relevant intrinsic damping. However, in spite of this disadvantage, we clearly stated that:

\begin{itemize}

\item the $\nu = \xi$ correctly determines a physical radial mass viscous transport in an annulus ring, as it is in the case of the Shakura and Sunyaev formulation. If the characteristic mixing length, determined by expressions like (17) or (23), is comparable with $L$, other formulations for $\nu \propto l$, or $\nu \propto l^{2}$ (eqs. 16, 17, 33) looks like too viscous, so much that the entire disc is quickly accreted.

\item the role of the numerical density, as well as both of the molecular-atomic cross section and of eddy linear dimension ($l$) on $\nu = \xi$ are evident in so far as the physical damping is at least comparable with that of the intrinsic dissipation. Eqs. (26 and 27) include all physical situations. The main contribution on $\nu = \xi$ either comes out from a long $l$, as in the case of the mass radial viscous transport in the annulus and in disc structures, or it comes out from a large numerical density, as in the case of random chaotic motions. Nevertheless, specifying the chemical composition of the fluid, what is relevant is not only $l$, as eqs. 26 and 27 show.

\item Expressions where $\nu \propto h$ are arbitrary, of course, despite low resolution results ofd simulations are physically meaningful, because there is not any physical correlation between the kinematic viscosity and the eddy linear size.

\item In the 3D accretion disc simulations, results on steady disc structure for $\nu = \xi$ are in a good accordance to those relative to the Shakura and Sunyaev formulation without any ambiguity on the arbitrary $\alpha_{SS}$ parameter. The comparison in both high and in low compressibility are very successful. In particular, for $\gamma = 1.01$, $\nu = \xi$ looks like working as $\nu = \alpha_{SS} c_{s} H$ with $\alpha_{SS} = 0.3 - 0.5$. Instead, other expressions for $\nu$ are unsuccessful either in one case or in the other case in so far as the Shakura and Sunyaev viscous prescription is correct. On the contrary, the highly viscous disc modelling could deeply pay attention to the relevance of the $\nu = c_{s} l$ kinematic viscosity coefficient in the solutions of the Navier-Stokes equations. It is true that high viscosity disc models could also be built up tuning either $\rho$ and/or $\overline{\kappa}$ for $\nu = \xi$. However, high viscosity physics could be obtained in high density conditions (up to terrestrial conditions, see eqs. 26, 27) that very rarely happens in diffuse matter astrophysical environments.

\end{itemize}

  For astrophysical objects, showing flat accretion disc structures, the $\nu = \alpha_{SS} c_{s} H$ Shakura and Sunyaev formulation is successfully adopted. In this case, $H \approx r_{disc} c_{s}/v_{\phi} \approx 10^{-2} r_{disc}$ for a correct application in a strict Keplerian tangential kinematics. $\alpha_{SS}$ ranges within $\sim 0.001 - 0.4$ according to the astrophysical object considered, not without any ambiguity. The only physical local component in this formulation is the sound velocity, that is assumed as a function of the mass accretor and of the radial distance only, in shockless conditions. It does not exist any consideration on why the Prandtl's $\nu \sim l^{2} \partial v/\partial x$ is not taken into account. In our formulation, every component of $\nu = \xi$ depends on the local conditions, with or without shock events. This, without any doubt complicates an exact calculation on $\nu$ since, apart the chemical composition, the entire $(\Sigma/\overline{\mu} m_{H}) l \overline{\kappa}$ or $(\rho/\overline{\mu} m_{H}) l^{2} \overline{\kappa}$ terms also depend on the local densities and on $l$, which could be very different from the local disc thickness $H$. This without any consideration on $\overline{\kappa}$ that could be very different from molecular-atomic scales to high temperature nuclear scales of cosmological interest. As an example, for AGN, the term $\overline{\kappa}/\overline{\mu} m_{H} \approx 0.1 - 1$ g$^{-1}$ cm$^{2}$ for protons in so far as $T \approx 10^{6} - 10^{7}$ K. This means that if we arbitrarily impose $l = H$, either $\Sigma$ or $\rho l \approx 10^{-2}$ g cm$^{-2}$. Being the first 2D $\Sigma$ value too high, instead the second 3D $\rho l$ value looks like much more plausible. For star forming objects like FU Orionis, the $\alpha_{SS} \approx 10^{-3}$ in the $\nu = \alpha_{SS} c_{s} H$ Shakura and Sunyaev formulation. If we consider a gas of pure neutral atomic hydrogen, the ratio $\overline{\kappa}/\overline{\mu} m_{H} \approx 10^{4}$ g$^{-1}$ cm$^{-2}$. Hence, $\Sigma$ or $\rho l \approx 10^{-12}$ g cm$^{-2}$, that means  a quite high value for $\Sigma$ or a low, but more realistic value for $\rho l$. The entire evaluation on $\nu = \xi$ so far includes molecular-atomic-nuclear characteristics only on the basis of a collisional molecular gas dynamics without any consideration on the role of electric and/or magnetic effects, as well as on the presence of dust in the diffuse matter. On the contrary, the evaluations of $\overline{\kappa}$ could be very complicated and its value very different, up to orders of magnitude toward much larger values. In this sense, we got the simplest and the lowest values for $\nu = \xi$ in this paper.

  Results shown throughout this paper were obtained working in the SPH framework. However, all results here discussed are not dependent on the adopted numerical code because physics of viscous dissipation is shown in its purely physical aspect, since we discussed a physical formulation for $\nu$. Parts of fluid, either as Lagrangian moving particles, or within Eulerian grid cells are conceptually the same both to the Euler and to the Navier-Stokes flow equations. In spite of fact that the artificial viscosity can also be tuned, according to the physical event considered,  being SPH still an intrinsically artificially "viscous" technique, the evidence of successful results here shown demonstrates that, even in the worst conditions, $\nu = \xi$ correctly works without any ambiguity, free of any arbitrary parameters, also including those molecular characteristics that are uncommon in other formulations.

\section*{Acknowledgments}

We thank Dr. S. Scuderi of the INAF - Osservatorio Astrofisico di Catania for some helpful interventions that improved the presentation of the paper.

\label{lastpage}


\begin{thebibliography}{99}

\bibitem[\protect\citeauthoryear{Abolmasov \& Shakura}{2009}]{c53} Abolmasov, P., Shakura, N.I., 2009, AN, 7, 737
\bibitem[\protect\citeauthoryear{Bate et al.}{2000}]{a36} Bate, M.R., Bonnell, I.A., Clarke, C.J., Lubow, S.H., Ogilvie, G.I., Pringle, J.E., Tout, C.A., 2000, MNRAS, 317, 773
\bibitem[\protect\citeauthoryear{Bell \& Lin}{1994}]{b6} Bell, K.R., Lin, D.N.C., 1994, ApJ, 427, 987
\bibitem[\protect\citeauthoryear{Bisikalo et al.}{1998a}]{a1} Bisikalo, D.V., Boyarchuk, A.A., Kutzetzov, O.A., 1998a, Astron. Rep., 42, 33
\bibitem[\protect\citeauthoryear{Bisikalo et al.}{1998b}]{a2} Bisikalo, D.V., Boyarchuk, A.A., Chechetkin, V.M., Kutzetzov, O.A., Molteni, D., 1998b, MNRAS, 300, 39
\bibitem[\protect\citeauthoryear{Bisikalo et al.}{1999}]{a3} Bisikalo, D.V., Boyarchuk, A.A., Chechetkin, V.M., Kutzetzov, O.A., 1999, Astron. Rep., 43, 797
\bibitem[\protect\citeauthoryear{Bisikalo et al.}{2000}]{a4} Bisikalo, D.V., Boyarchuk, A.A., Kutzetzov, O.A., Chechetkin, V.M., 2000, Astron. Rep., 44, 26
\bibitem[\protect\citeauthoryear{Clarke et al.}{1990}]{b5} Clarke, C.J., Lin, D.N.C., Pringle, J.E., 1990, MNRAS, 242, 439
\bibitem[\protect\citeauthoryear{Fletcher}{1991}]{c3} Fletcher, C.A.J., 1991, "Computational techniques for fluid dynamics", Springer
\bibitem[\protect\citeauthoryear{Frank et al.}{2002}]{a69} Frank, J., King, A., Raine, D., 2002, "Accretion Power in Astrophysics", Cambridge Univ. Press
\bibitem[\protect\citeauthoryear{Frish}{1995}]{a58} Frish, U., 1995, "Turbulence", Cambridge Univ. Press
\bibitem[\protect\citeauthoryear{Hartman et al.}{1998}]{b4} Hartman, L., Calvet, N., Gullbring, E., D'Alessio, P., 1998, ApJ, 495, 385
\bibitem[\protect\citeauthoryear{Hirsch}{1997}]{c4} Hirsch, C., 1997, "Numerical computation of internal and external flows", Wiley
\bibitem[\protect\citeauthoryear{Ichikawa \& Osaki}{1992}]{a8} Ichikawa, S., Osaki, Y., 1992, PASJ, 44, 15
\bibitem[\protect\citeauthoryear{Ichikawa \& Osaki}{1994}]{a10} Ichikawa, S., Osaki, Y., 1994, PASJ, 46, 621
\bibitem[\protect\citeauthoryear{Katz}{1973}]{a34} Katz, J.I., 1973, Nature Phys. Sci., 246, 87
\bibitem[\protect\citeauthoryear{Kellay \& Goldburg}{2002}]{a59} Kellay, H., Goldburg, W.I., 2002, Rep. Prog. Phys., 65, 845
\bibitem[\protect\citeauthoryear{King et al.}{2007}]{c52} King, A.R., Pringle, J.E., Livio, M., 2007, MNRAS, 376, 1740
\bibitem[\protect\citeauthoryear{Kley et al. et al.}{2008}]{a33} Kley, W., Papaloizou, J.C.B., Ogilvie, G.I., 2008, A\&A, 487, 671
\bibitem[\protect\citeauthoryear{Kolmogorov}{1941a}]{a72} Kolmogorov, A.N., 1941a, "The local structure of turbulence in incompressible viscous fluid for very large Reynolds numbers". Proc. of the USSR Academy of Sciences 30, p.299303. (Russian), translated into English by Kolmogorov, A.N., 1991. "The local structure of turbulence in incompressible viscous fluid for very large Reynolds numbers". Proc. of the Royal Society of London, Series A: Math. and Phys. Sci. 434, 9-13.
\bibitem[\protect\citeauthoryear{Kolmogorov}{1941b}]{a73} Kolmogorov, A.N., 1941b, Kolmogorov, "Dissipation of energy in locally isotropic turbulence". Proc. of the USSR Academy of Sciences 32, p.16-18. (Russian), translated into English by Kolmogorov, A.N., 1991. "The local structure of turbulence in incompressible viscous fluid for very large Reynolds numbers". Proc. of the Royal Society of London, Series A: Math. and Phys. Sci. 434, p.15-17.
\bibitem[\protect\citeauthoryear{Lanzafame}{2003}]{a5} Lanzafame, G. 2003, A\&A, 403, 593
\bibitem[\protect\citeauthoryear{Lanzafame}{2008}]{c45} Lanzafame, G. 2008, PASJ, 60, 259
\bibitem[\protect\citeauthoryear{Lanzafame}{2009}]{c46} Lanzafame, G., 2009, AN, 330, 843
\bibitem[\protect\citeauthoryear{Lanzafame}{2010a}]{c16} Lanzafame, G., 2010a, MNRAS, 408, 2336
\bibitem[\protect\citeauthoryear{Lanzafame}{2010b}]{c24} Lanzafame, G., 2010b, ASP, 429, 106
\bibitem[\protect\citeauthoryear{Lanzafame et al.}{1992}]{a38} Lanzafame, G., Belvedere G., Molteni D., 1992, MNRAS, 258, 152
\bibitem[\protect\citeauthoryear{Lanzafame et al.}{1998}]{b22} Lanzafame, G., Molteni, D., Chakrabarti, S.K., 1998, MNRAS, 299, 799
\bibitem[\protect\citeauthoryear{Lanzafame et al.}{2000}]{c47} Lanzafame, G., Maravigna, F., Belvedere G., 2000, PASJ, 52, 515
\bibitem[\protect\citeauthoryear{Lanzafame et al.}{2001}]{c59} Lanzafame, G., Maravigna, F., Belvedere G., 2001, PASJ, 53, 139
\bibitem[\protect\citeauthoryear{Lanzafame et al.}{2008}]{b23} Lanzafame, G., Cassaro, P., Schillir\'o, F., Costa, V., Belvedere, G., Zappal\'a, R.A., 2008, A\&A, 482, 473
\bibitem[\protect\citeauthoryear{Lanzafame et al.}{2011}]{c25} Lanzafame, G., Costa, V., Belvedere, G., 2011, ASP, 444, 230
\bibitem[\protect\citeauthoryear{LeVeque}{1992}]{c37} LeVeque, R.J., 1992, "Numerical methods for conservation laws", Lectures in Mathematics, ETH Z\"urich, Birkh\"auser
\bibitem[\protect\citeauthoryear{LeVeque}{2002}]{c5} LeVeque, R.J., 2002, "Finite volume methods for hiperbolic problems", Cambridge Univ. Press
\bibitem[\protect\citeauthoryear{Lodato \& Clarke}{2004}]{b7} Lodato, G., Clarke, C.J., 2004, MNRAS, 353, 841
\bibitem[\protect\citeauthoryear{Lubow \& Shu}{1975}]{c49} Lubow, S.H., Shu, F.H.,  1975, MNRAS, 198, 383
\bibitem[\protect\citeauthoryear{Manz et al.}{2009}]{a63} Manz, P., Ramisch, M., Stroth, U., 2009, Plasma Phys. and Controlled Fusion, 51, 35008
\bibitem[\protect\citeauthoryear{Molteni et al.}{1991}]{c48} Molteni, D., Belvedere, G., Lanzafame, G., 1991, MNRAS, 249, 748
\bibitem[\protect\citeauthoryear{Murray}{1996}]{a9} Murray, J.R., 1996, MNRAS, 279, 402
\bibitem[\protect\citeauthoryear{Ogilvie \& Dubus}{2000}]{a37} Ogilvie, G.I., Dubus, G., 2000, MNRAS 320, 485
\bibitem[\protect\citeauthoryear{Okazaki et al.}{2002}]{a11} Okazaki, A.T., Bate, M.R., Ogilvie, G.I., Pringle, J.E., 2002, MNRAS 337, 967
\bibitem[\protect\citeauthoryear{Papaloizou \& Pringle}{1977}]{a6} Papaloizou, J.C.B., Pringle, J.E., 1977, MNRAS, 181, 441
\bibitem[\protect\citeauthoryear{Park \& Kwon}{2003}]{a19} Park, S.H., Kwon, J.H., 2003, JCoPh, 188, 524
\bibitem[\protect\citeauthoryear{Prandtl}{1925}]{b1} Prandtl, L., 1925, Math. Mech, 5, 136
\bibitem[\protect\citeauthoryear{Pringle}{1981}]{b3} Pringle, J.E., 1981, ARA\&A, 19, 137
\bibitem[\protect\citeauthoryear[Reif (1965)]{a39} Reif, F. 1965, "Fundamentals of Statistical and Thermal Physics", McGraw-Hill Co.
\bibitem[\protect\citeauthoryear{Saffman}{1971}]{a82} Saffman, P.G., 1971, Stud. Appl. Math., 377, 383
\bibitem[\protect\citeauthoryear{Sawada et al.}{1987}]{b8} Sawada, K., Matsuda, T., Inoue, M., Hachisu, I., 1987, MNRAS, 224, 307
\bibitem[\protect\citeauthoryear{Schmitt}{2007}]{b10} Schmitt, F., 2007, Compt. Rend. Mec., 335, 617
\bibitem[\protect\citeauthoryear{Shakura}{1972}]{c50} Shakura, N.I. 1972, Astron. Zh., 49, 921. (English tr.: 1973, Sov. Astron., 16, 756)
\bibitem[\protect\citeauthoryear{Shakura \& Sunyaev}{1973}]{c51} Shakura, N.I., Sunyaev, R.A., 1973, A\&A, 24, 337
\bibitem[\protect\citeauthoryear{Speith \& Kley}{2003}]{b11} Speith, R., Kley, W., 2003, A\&A, 399, 395
\bibitem[\protect\citeauthoryear{Spruit et al.}{1987}]{b9} Spruit, H.C., Matsuda, T., Inoue, M., Sawada, K., 1987, MNRAS, 229, 517
\bibitem[\protect\citeauthoryear{Tabeling}{2002}]{a67} Tabeling, P., 2002, Phys. Rep., 362, 1
\bibitem[\protect\citeauthoryear{Toro}{1999}]{a71} Toro, E.G., 1999, "Riemann solvers and numerical methods for fluid dynamics", Springer-Verlag
\bibitem[\protect\citeauthoryear{Trampedach \& Stein}{2011}]{b2} Trampedach, R., Stein, R.F., 2011, ApJ, 731, 78
\bibitem[\protect\citeauthoryear{Whitehurst}{1988a}]{a20} Whitehurst, R., 1988, MNRAS, 232, 35
\bibitem[\protect\citeauthoryear{Whitehurst}{1988b}]{a21} Whitehurst, R., 1988, MNRAS, 233, 529
\bibitem[\protect\citeauthoryear{Wijers \& Pringle}{1999}]{a35} Wijers, R.A.M.J., Pringle, J.E., 1999, MNRAS, 308, 207
\bibitem[\protect\citeauthoryear{Zang \& Chen}{1992}]{a7} Zhang, Z.Y., Chen, J.S., 1992, A\&A, 261, 493

\end{thebibliography}
\end{document}